\documentclass[sigconf]{acmart}

\renewcommand\footnotetextcopyrightpermission[1]{} % removes footnote with conference information in first column
\usepackage{booktabs,amsmath,amssymb,subcaption,hyperref,listings,color,braket,multicol}
\hypersetup{breaklinks=true}
\graphicspath{{./figures/}}

\newcommand{\tket}{t$|$ket$\rangle$}

\begin{document}

%%
%% The "title" command has an optional parameter,
%% allowing the author to define a "short title" to be used in page headers.
\title{Using Reinforcement Learning to Perform Qubit Routing in Quantum Compilers}

%%
%% The "author" command and its associated commands are used to define
%% the authors and their affiliations.
%% Of note is the shared affiliation of the first two authors, and the
%% "authornote" and "authornotemark" commands
%% used to denote shared contribution to the research.
\author{Matteo G. Pozzi}
\affiliation{%
  \institution{University of Cambridge Computer Laboratory}
}
\email{mgp35@cam.ac.uk}

\author{Steven J. Herbert}
\affiliation{%
  \institution{University of Cambridge Computer Laboratory}
}
\affiliation{%
  \institution{Cambridge Quantum Computing}
}
\email{sjh227@cam.ac.uk}

\author{Akash Sengupta}
\affiliation{%
  \institution{Department of Engineering, University of Cambridge}
}
\email{as2562@cam.ac.uk}

\author{Robert D. Mullins}
\affiliation{%
  \institution{University of Cambridge Computer Laboratory}
}
\email{rdm34@cam.ac.uk}

%%
%% By default, the full list of authors will be used in the page
%% headers. Often, this list is too long, and will overlap
%% other information printed in the page headers. This command allows
%% the author to define a more concise list
%% of authors' names for this purpose.
\renewcommand{\shortauthors}{Pozzi et al.}

%%
%% The abstract is a short summary of the work to be presented in the
%% article.
\begin{abstract}
``Qubit routing'' refers to the task of modifying quantum circuits so that they satisfy the connectivity constraints of a target quantum computer. This involves inserting SWAP gates into the circuit so that the logical gates only ever occur between adjacent physical qubits. The goal is to minimise the circuit depth added by the SWAP gates.

In this paper, we propose a qubit routing procedure that uses a modified version of the deep Q-learning paradigm. The system is able to outperform the qubit routing procedures from two of the most advanced quantum compilers currently available, on both random and realistic circuits, across near-term architecture sizes.
\end{abstract}

\maketitle

\section{Introduction}

%% Introduction

In his highly influential 2018 paper, John Preskill coined the term \textit{Noisy Intermediate-Scale Quantum} (NISQ) technology \cite{Preskill2018}, and suggested that the so-called ``NISQ era'' would arrive in the near future: that is, we would soon have quantum computers with 50-100 qubits that would be able to solve problems that are intractable for even the best classical computers (a phenomenon known as \textit{quantum supermacy} \cite{Preskill2012}). Since then, we have seen Google \cite{Martinis2019} and more recently IBM \cite{Chow2020} produce quantum computers with over 50 qubits. In October 2019, Google announced they had achieved quantum supremacy with their 53-qubit Sycamore processor \cite{Arute2019} (although IBM were quick to dispute this claim \cite{Pednault}).

Such near-term quantum computers support a series of one- and two-qubit operations, or ``gates'', which can be assembled into quantum ``circuits'', in a similar spirit to how logic gates that act on classical bits can be assembled into sequential logic circuits. A high-level circuit description (in a language such as OpenQASM \cite{Cross2017}) must be compiled before it can be executed on a target quantum architecture --- this process includes passes to satisfy the constraints of the target hardware. Specifically, each quantum architecture has an associated ``topology'', or connectivity graph, consisting of a set of physical nodes and links between them. Qubits inhabit the nodes and can only interact with qubits on adjacent nodes --- SWAP gates can swap the nodes they inhabit.

To make an arbitrary quantum circuit executable on a given target architecture, a quantum compiler has to insert SWAP gates so that gates in the original circuit only ever occur between qubits located at adjacent nodes, a process known as ``routing''. This will produce a new circuit, possibly with a greater depth, that implements the same unitary function as the original circuit while respecting the topological constraints.

The quantum architectures of today are extremely resource-constrained devices, with relatively low numbers and fidelities of qubits. Minimising the added circuit depth is a key goal in maximising the amount of useful work that can be done by today's systems before decoherence, so much so that in 2018, IBM offered a prize for the best qubit routing algorithm \cite{IBM_Contest_Winners}. Tan and Cong \cite{Tan2020} recently compared the performance of various state-of-the-art routing algorithms on benchmarks with known optimal depth, and concluded that even the most advanced algorithms are significantly lacking --- scope therefore exists to improve upon them.

% Wording should be ok now

In this paper, we frame the qubit routing problem as a reinforcement learning (RL) problem, building on previous work by Herbert and Sengupta \cite{Herbert2018}, which employed a modified deep Q-learning approach to route qubits. While the work certainly demonstrated that such an approach had potential, its RL formulation did not fully address the concerns of a real quantum compiler, most notably perhaps the fact that quantum gates take some time to execute. In this work, we make numerous improvements to the original system and propose a new RL formulation that addresses these concerns, such that the new system now represents a solution to the qubit routing problem. We then benchmark the system against the routing passes of state-of-the-art quantum compilers, and demonstrate that our system is able to outperform its competitors on the most pertinent benchmarks.

\section{Background}

%% Background

In this section, we begin by formalising and defining the terms used throughout the paper. We then outline the RL approach proposed by Herbert and Sengupta \cite{Herbert2018}, since this work builds upon it directly.

\subsection{Quantum circuits}

\begin{figure}
\centering
\begin{subfigure}[c]{.4\linewidth}
  \centering
  \includegraphics[height=0.8\linewidth]{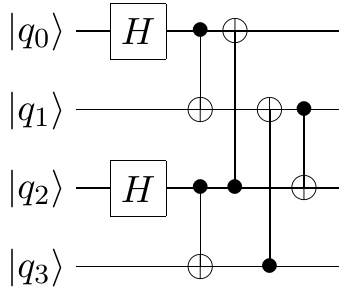}
  \caption{Quantum circuit}
  \label{fig:example_circuit_1}
\end{subfigure}
\hspace{5mm}
\begin{subfigure}[c]{.4\linewidth}
  \centering
  \includegraphics[height=0.8\linewidth]{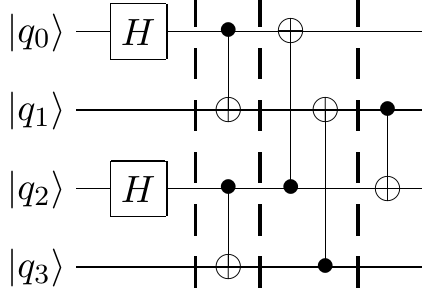}
  \caption{Its decomposition}
  \label{fig:example_circuit_1_with_layers}
\end{subfigure}
\caption{An example of a quantum circuit and its decomposition into layers}
\label{fig:background_example_circuit}
\end{figure}

A quantum circuit is composed of a series of operations, or gates, which transform the state of one or more logical qubits. Figure \ref{fig:example_circuit_1} shows an example of a quantum circuit with four qubits --- this circuit contains two Hadamard gates and five Controlled-NOT (CNOT) gates in various orientations.

Quantum circuits can be decomposed into a universal set of one- and two-qubit gates --- on many architectures, the two-qubit gate of choice is the CNOT. In this paper we therefore only consider CNOT gates --- single-qubit gates are not relevant in qubit routing since they can occur at any node, and in any case, they are much quicker than two-qubit gates on all real quantum computers.

% TODO: numbering of qubits here?

An important concept is the notion of circuit \textit{depth}. For an ordered set of two-qubit gates $G$, each acting on qubits $q_j$ and $q_k$ (such that $g_i = \{q_j, q_k\}$), indexed from $i$ = 1:

\begin{equation}
d_0(q) = 0 \text{ for all qubits } q \text{ in the circuit}
\end{equation}

\begin{equation}
d_{t+1}(q) = \begin{cases} max(d_t(q_j), d_t(q_k)) + 1 & \mbox{for } q_j, q_k \in g_t \mbox{ if } q \in g_t \\ 
d_t(q) & \mbox{otherwise} \end{cases}
\end{equation}

\begin{equation}
\text{Circuit depth } d = \max_{q}(d_{|G|}(q))
\end{equation}

This can be visualised as slicing the circuit into \textit{timesteps} of gates that can be performed in parallel --- the depth of a circuit is then the minimum number of timesteps it can be decomposed into, without any qubit performing more than one interaction in any given timestep (see Figure \ref{fig:example_circuit_1_with_layers}).

\subsection{Quantum architectures}

\begin{figure}
  \centering
  \includegraphics[width=0.6\linewidth]{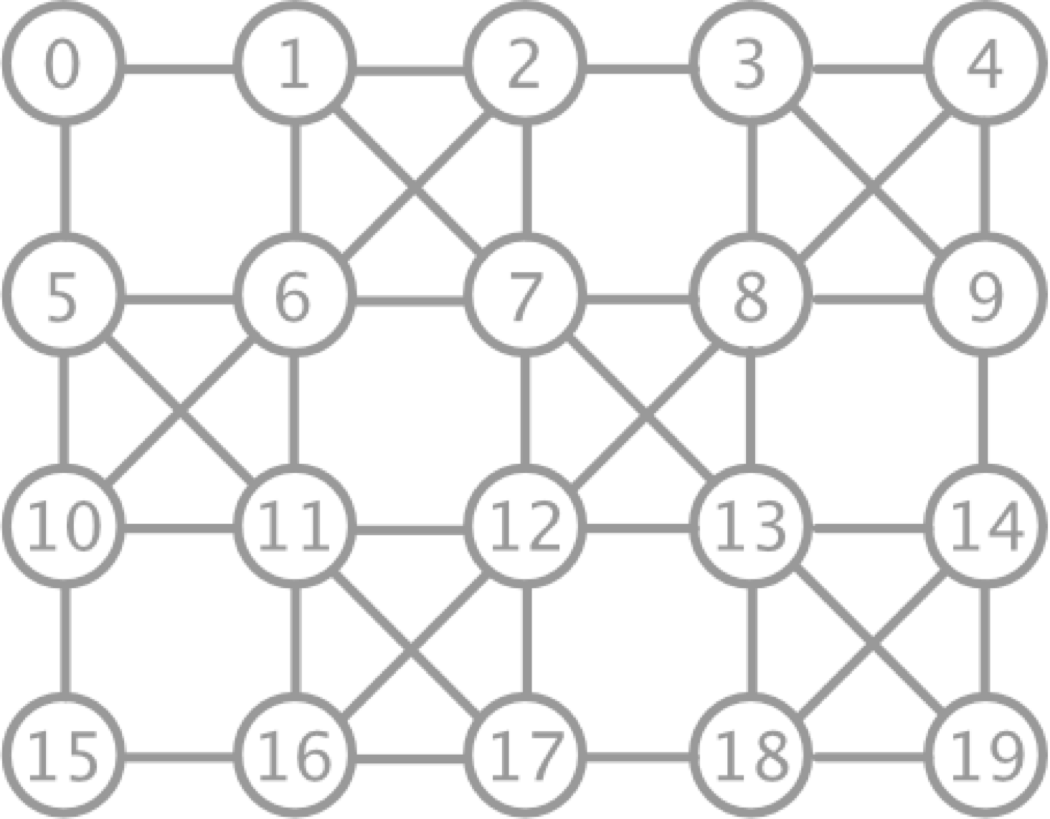}
  \caption{The IBM Q20 Tokyo \cite{IBM_Devices}}
  \label{fig:ibm_q20}
\end{figure}

For our purposes, a quantum architecture is a connectivity graph, composed of a set of physical qubits, or \textit{nodes}, and a set of links between them. Figure \ref{fig:ibm_q20} provides an example of one of IBM's quantum architectures with 20 qubits. In this paper, we only consider \textit{undirected} qubit connectivity graphs --- that is, CNOT gates can be performed in either direction. In practice, the direction of CNOT gates can be inverted by using Hadamard gates if necessary, so this simplification makes little difference in our domain.

\subsection{Placements and SWAP gates}

At any time $t$ during circuit execution, qubits $Q$ are mapped onto nodes $N$ according to some placement \(p_t: Q \rightarrow N\). Gates may only occur between two qubits if they lie on adjacent nodes --- that is, gate \(g = \{q_0, q_1\}\) may only occur at time $t$ if \((p_t(q_0), p_t(q_1)) \in E\), where $E$ is the set of edges in the architecture's connectivity graph.

\begin{figure}
\centering
\begin{subfigure}[c]{.3\linewidth}
  \centering
  \includegraphics[width=.9\linewidth]{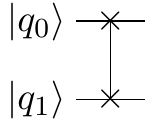}
  \caption{A SWAP gate}
  \label{fig:swap_gate}
\end{subfigure}
=
\begin{subfigure}[c]{.4\linewidth}
  \centering
  \includegraphics[width=.9\linewidth]{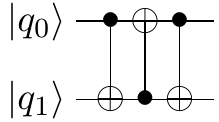}
  \caption{Three CNOT gates}
  \label{fig:swap_gate_cnot}
\end{subfigure}
\caption{A SWAP gate and its decomposition into three CNOT gates}
\label{fig:swap_gates}
\end{figure}

SWAP gates allow two qubits on adjacent nodes to switch positions. Formally, for SWAP gate \(s = \{q_0, q_1\}\) at time $t$, \(p_{t+1}(q_0) = p_t(q_1)\) and vice versa. Figure \ref{fig:swap_gates} shows the circuit symbol for a SWAP gate, which can be decomposed into three CNOT gates in sequence.

% TODO: mention same time assumption?

\subsection{Qubit routing}

\begin{figure*}
\centering
\begin{subfigure}{.2\linewidth}
  \centering
  \includegraphics[height=0.9\linewidth]{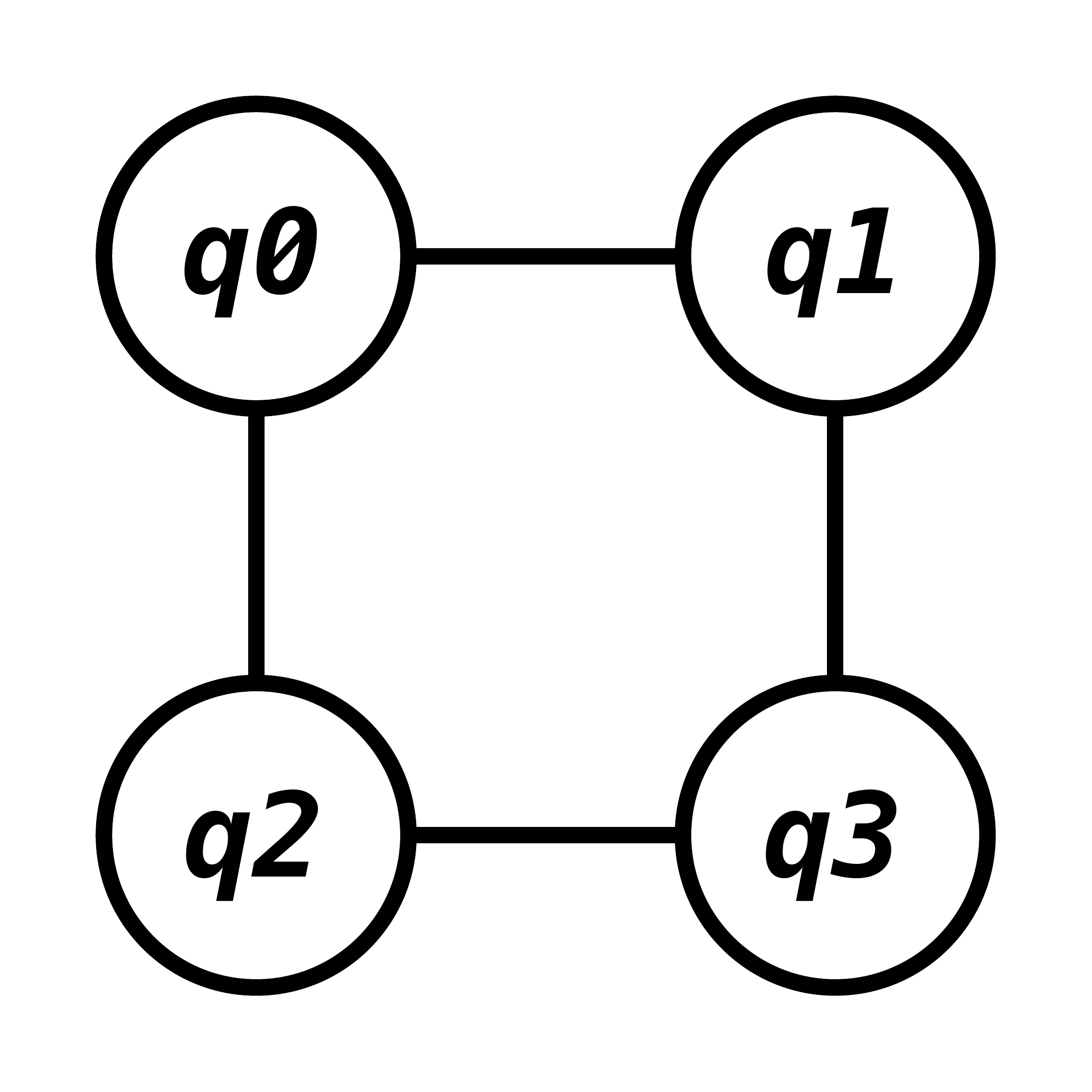}
  \caption{4-qubit topology}
  \label{fig:2x2_grid}
\end{subfigure}
\hspace{2mm}
\begin{subfigure}{.3\linewidth}
  \includegraphics[height=0.6\linewidth]{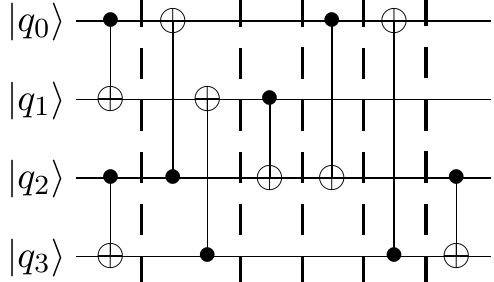}
  \caption{Before routing}
  \label{fig:example_circuit_3}
\end{subfigure}
\hspace{5mm}
\begin{subfigure}{.4\linewidth}
  \includegraphics[height=0.45\linewidth]{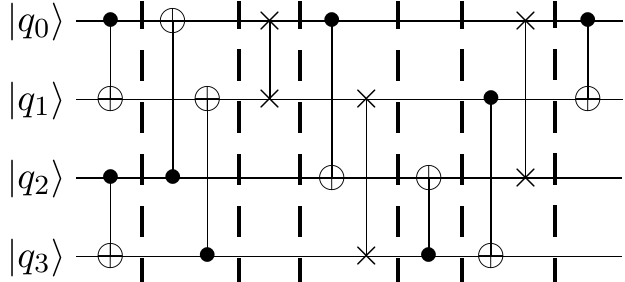}
  \caption{After routing}
  \label{fig:example_circuit_3_routed}
\end{subfigure}
\caption{An example of a quantum circuit before and after the routing procedure}
\label{fig:evaluation_example_circuit}
\end{figure*}

``Routing'' denotes the task of inserting SWAP gates into quantum circuits so that every gate in the original circuit can be performed on a given target architecture. Qubit routing passes in quantum compilers generally accept a circuit together with a connectivity graph and initial layout, and output a new circuit that respects these architectural constraints. The routing process can thus be represented as a function $R: (c,g,l) \mapsto c'$, with input circuit $c$, output circuit $c'$, connectivity graph $g$, and initial layout $l$.

The goal is to minimise the added depth of the output circuit versus that of the original circuit. This goal is important because we want to try and maximise the amount of useful work performed by qubits before decoherence, and thus maximise the Quantum Volume (QV) \cite{Chow2020} our current systems can achieve. Added depth can be formalised as two metrics, \textit{circuit depth overhead} (CDO) and \textit{circuit depth ratio} (CDR):

\begin{equation}
CDO = d(c') - d(c) \quad \quad CDR = \frac{d(c')}{d(c)}
\end{equation}

\noindent where $d$ denotes circuit depth, $c$ is the original circuit, and $c'$ is the routed circuit.

Figure \ref{fig:evaluation_example_circuit} shows a quantum circuit of depth 6, before and after routing on the topology and initial layout in Figure \ref{fig:2x2_grid}. The routed circuit has depth 7, and therefore a CDO of 1 and a CDR of $\frac{7}{6}$. Notice how two of the SWAP gates do not add any extra depth to the circuit --- the routing procedure is able to perform these while CNOT gates are happening, thus minimising the depth overhead.

The process of routing qubits is inherently linked to their initial placement $p_0$. In this work we mostly consider random initial placements, for fair comparisons of the routing algorithms themselves, although many quantum compilers do provide strategies to optimise initial placement and these are also important (for some simple circuits, inserting swaps might not be necessary at all).

\subsection{RL formulation for qubit routing}

%% The Original Approach

Reinforcement Learning (RL) is a sub-field of Machine Learning that offers a powerful paradigm for learning to perform tasks in contexts with very little \textit{a priori} knowledge of what the optimal strategy might be. It has proven to be useful in complex situations with lots of input data, such as robotics \cite{Kober2013} and video games \cite{Mnih2013,Karnsund2019}. Under the paradigm, agents learn to achieve some goal in an environment by freely interacting with it and observing rewards for performing actions in different states (see Figure \ref{fig:rl_loop}). % changed

In this section we outline the RL formulation for qubit routing proposed by Herbert and Sengupta \cite{Herbert2018}. Their goal was to frame the qubit routing problem as a reinforcement learning problem, employing a modified version of the Deep Q-learning (DQN) paradigm \cite{Mnih2013} --- the original version consists of a convolutional neural network to learn a function $Q(s,a)$ that represents the quality of being in state $s$ and taking action $a$. A description of the modified formulation is as follows:

\begin{figure}
\centering
\includegraphics[width=0.95\linewidth]{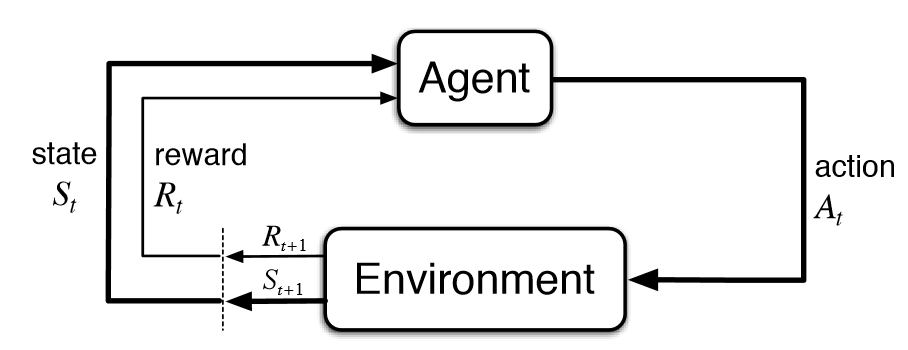}
  \caption{An illustration of the RL process \cite{RL_Loop}}
  \label{fig:rl_loop}
\end{figure}

\begin{itemize}
  \item \textbf{State}: a tuple encoding information about the current state of the architecture and progress through the circuit
    \begin{itemize}
      \item \textit{Qubit locations}: a mapping $l: N \rightarrow Q$ denoting which architectural nodes are currently holding which logical qubits.
      \item \textit{Qubit targets}: a partial mapping $t: Q \rightharpoonup Q$, with $q_2 = t(q_1)$ iff $q_1$'s next interaction is with $q_2$, or undefined if $q_1$ has performed all of its interactions. % changed
      \item \textit{Circuit progress}: a mapping $p: Q \rightarrow \mathbb{N}_{d+1}$, for a circuit with depth $d$, with $n = p(q)$ iff $q$ has completed $n$ interactions so far.
    \end{itemize}
  \item \textbf{Reward}: the main reward signal is issued when scheduling (i.e.\ performing) a gate, but other reward signals are also used (see below)
  \item \textbf{Action}: an action is a set of swaps $S$ that can be performed in parallel --- logical gates are performed implicitly % changed
\end{itemize}

The key insight lies in the fact that the action space is combinatorial --- that is, for a connectivity graph with $n$ edges, there are $O(2^n)$ possible parallelisable sets of SWAP gates to choose from. It is therefore intractable to have a neural network learn a quality function based on state-action pairs. Instead, it learns the quality of individual states, and then a combinatorial optimisation technique (simulated annealing in this case) can be used to search for the action leading to the highest-quality next state. The quality function is therefore as follows:

\begin{equation}
Q(s) = r^*(s) + \gamma \max_{a} Q(\text{env}(s,a))
\end{equation}

with $r^*$ yielding the reward available from an optimal action selection policy, and the env function producing the next state for a given state-action pair. This modified Q-learning process can then be summarised as follows:

\begin{equation}
Q(s_t) \leftarrow (1 - \alpha) Q(s_t) + \alpha \left(r_t + \gamma \max_{a_{t+1}} Q(s_{t+1})\right)
\end{equation}

\noindent for observed experience tuple $(s_t, a_t, r_t, s_{t+1})$ and learning rate $\alpha$. The above state representation and action space form a fully deterministic Markov Decision Process (MDP) --- given the current state of the system, applying a given action will always produce the same next state.

A feature selection function $\phi$ is used to condense the above state format into a more succinct vector representation to be passed to the neural network --- this representation is effectively a (partial) mapping $t': N \rightharpoonup N$ that maps nodes to their targets (partial in the sense that some nodes may have no target, namely those that hold qubits that do not participate in any further gates). Formally, for node $n_1$ with target $n_2$, $t': n_1 \mapsto n_2$ iff $l(n_1) = q_1 \wedge l(n_2) = q_2 \wedge t(q_1) = q_2$. In terms of function composition, $t': n \mapsto (l^{-1} \circ t \circ l)(n)$. This is essential to ensure that the input state to the neural network is always of a fixed size, and crucially, not too large.

The learning process makes use of two improvements to deep Q-learning, Double Deep Q-learning (DDQN) \cite{VanHasselt2016} and Prioritised Experience Replay (PER) \cite{Schaul2016}. The former helps improve the stability of the learning process by using two neural networks instead of one, while the latter enhances the learning process by replaying more useful experiences with a higher priority.

\subsubsection{Performing the actions}

In standard RL style, the environment takes a state $s_t$ and action $a_t$, and outputs a tuple $(s_{t+1},r_t)$, as follows:

\begin{enumerate}
  \item Calculate the total distance between each pair of mutually-targeting nodes (i.e.\ gates) --- call this the total pre-swap distance $d_{pre}$
  \item Perform the swaps in $a_t$, and calculate the total post-swap distance $d_{post}$
  \item ``Schedule'' the gates to produce state $s_{t+1}$ --- that is, perform any gates whose nodes are a distance of one away from each other
  \item Compute reward $r_t$
\end{enumerate}

A fixed reward is offered for each gate scheduled; if none are scheduled, a reward is offered based on whether $d_{post} < d_{pre}$, i.e.\ whether the action has brought qubits closer to their targets on average. A circuit completion reward is also offered. The key point to note is that once mutually-targeting nodes land a distance of one away from each other, the gate is performed immediately, taking zero timesteps. The initial state is created by generating a random initial placement of qubits, and reducing any gates that can be immediately performed.

\subsubsection{Simulated annealing}

This is the combinatorial optimisation process used to find actions to perform. The process searches for higher quality states by first swapping a random edge in the architecture, and then probing neighbouring solutions (i.e.\ actions that are one further edge swap away) and accepting them based on some acceptance probability. Actions that would lead to a non-parallelisable set of SWAPs are immediately disqualified and removed from further consideration.

The quality of a given action $a$, which acts on state $s_0$ to yield next state $s_1$ is just $Q(\phi(s_1))$. The acceptance probability is then:

\begin{equation}
P_{acc}(Q_0,Q_1,t) = \begin{cases} e^{(Q_1 - Q_0)/t} & \mbox{if } Q_1 \leq Q_0 \\ 
1 & \mbox{otherwise} \end{cases}
\end{equation}

\noindent for qualities $Q_0 = Q(\phi(s_0))$ and $Q_1 = Q(\phi(s_1))$, and current ``temperature'' $t$. The temperature decays by a fixed multiplier upon each iteration, until a given minimum temperature is reached.

\subsubsection{Static heuristics}

Random actions are taken as part of an $\epsilon$-greedy exploration policy, with a value of $\epsilon$ that begins at 1 and gradually decays after each learning batch. This approach suffers from some convergence issues, likely due to the fact that assigning a quality only to states (and not actions) allows e.g.\ a random action to tarnish the quality of an otherwise high-quality state. To tackle this issue, the authors include some static heuristics in their approach, such as forcing the movement of qubits in certain situations.

\section{Related Work}

%% Related Work

\subsection{Challenges of the NISQ era}

Quantum Computing in the NISQ era presents several challenges. Gyongyosi and Imre \cite{Gyongyosi2019} provide a recent summary of the space. Martonosi and Roetteler \cite{Martonosi2019} give a broad overview of the importance of computer science principles in the development of quantum computing. M\"oller and Vuik \cite{Moller2017} describe the lessons that can be learned from how classical architectures have evolved to do scientific computing. Almud\'ever et al. \cite{Almudever2017} provide a summary from an engineering perspective.

The general theme among these papers is that quantum computing is currently facing many of the challenges that classical computing faced during is development. Franke et al. \cite{Franke2019} highlight an important concern that came about in the late 1950's, called ``the tyranny of numbers'', which refers to the massive increase in the number of components and interconnections required as architectures were scaled up, before the invention of the integrated circuit. The authors note that current quantum computing architectures will see a similar lack of scalability --- they adapt Rent's rule \cite{Landman1971} and define a quantum Rent exponent $p$ to quantify the progress made in this aspect of optimisation.

Another key challenge is the lack of qubit fidelity, causing limits on achievable Quantum Volume (QV) --- this is a hardware-agnostic metric coined by IBM \cite{Chow2020}, which quantifies the limits of executable circuit size on a given quantum device. Minimising circuit depth provides an effective way of maximising QV, which is why the problem of routing qubits is such a key piece of the puzzle when it comes to maximising the usefulness of NISQ-era machines.

\subsection{Qubit routing}

\begin{figure*}
\centering
\begin{subfigure}[c]{.18\linewidth}
  \centering
  \includegraphics[height=\linewidth]{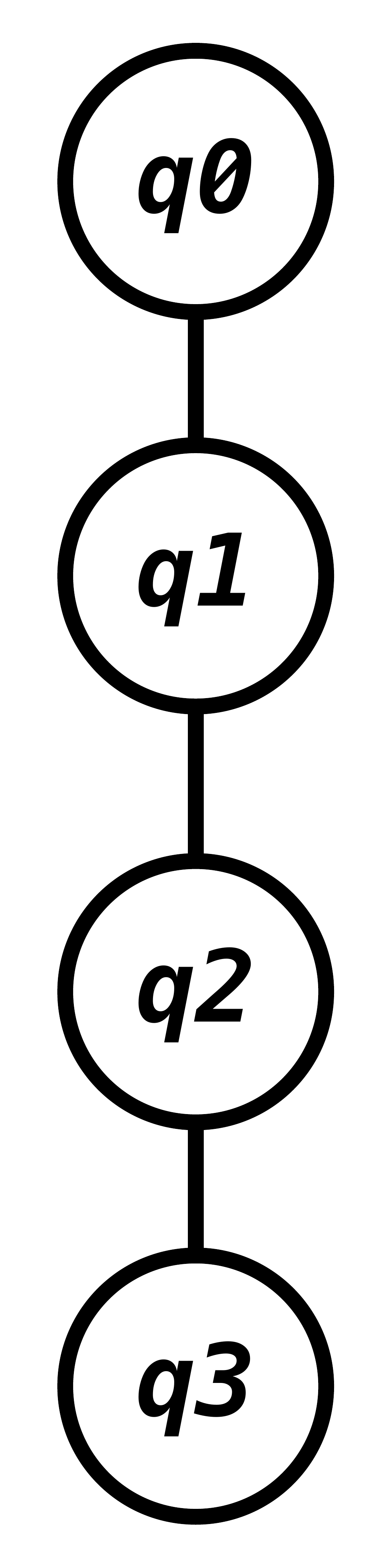}
  \caption{4-qubit topology}
  \label{fig:4x1_line}
\end{subfigure}
\begin{subfigure}[c]{.18\linewidth}
  \centering
  \includegraphics[height=\linewidth]{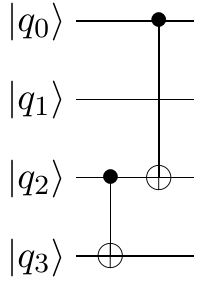}
  \caption{Circuit}
  \label{fig:new_rl_3}
\end{subfigure}
\begin{subfigure}[c]{.18\linewidth}
  \centering
  \includegraphics[height=\linewidth]{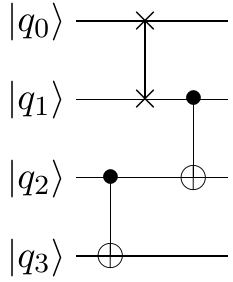}
  \caption{Routed circuit $c_1$}
  \label{fig:new_rl_1}
\end{subfigure}
\begin{subfigure}[c]{.18\linewidth}
  \centering
  \includegraphics[height=\linewidth]{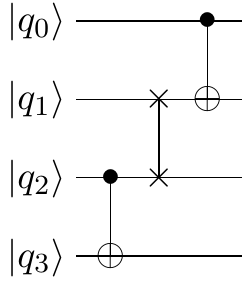}
  \caption{Routed circuit $c_2$}
  \label{fig:new_rl_2}
\end{subfigure}
\caption{A circuit and target topology, and two possible routed circuits.}
\label{fig:new_rl_example}
\end{figure*}

Herbert \cite{Herbert2018a} provides some theoretical bounds on the depth overhead incurred from routing, while Tan and Cong \cite{Tan2020} provide insight into the performance of various state-of-the-art routing systems on benchmarks with known optimal depth.

IBM's Qiskit \cite{IBM_Qiskit} is considered to be the most advanced and complete open-source quantum compiler. In 2018, IBM offered a prize for whoever could write the best routing algorithm for their quantum architectures \cite{IBM_Contest}. The winner \cite{IBM_Contest_Winners} was Alwin Zulehner, with an algorithm based on A* search \cite{Zulehner2019}. Second place was tied between Sven Jandura \cite{Jandura} and Eddie Schoute \cite{Schoute,Childs2019}. Zulehner claims that his background in Computer-Aided Design (CAD) helped guide his prize-winning approach \cite{Zulehner}, which further supports the theme of Section 3.1 (above). Zulehner has gone on to publish an approach targeting \textit{SU(4)} quantum circuits \cite{Zulehner2018}, and an approach for arbitrary quantum circuits based on Boolean satisfiability (SAT) \cite{Wille2019}, which is only tractable for circuits with small numbers of qubits. IBM have currently only integrated Jandura's approach into Qiskit \cite{LookaheadSwap}, although open-source code exists for the others \cite{ZulehnerRouting,SchouteRouting}.

Cambridge Quantum Computing's (CQC) \tket\ \cite{CQC_Technology} is a proprietary compiler with state-of-the-art performance. Its routing procedure is detailed by Cowtan et al. \cite{Cowtan2019} --- it effectively adds SWAPs to minimise some cost function based on proprietary heuristics. The \tket\ documentation suggests that the system now uses BRIDGE gates --- Itoko et al. \cite{Itoko2019} provide some insight into how the use of BRIDGE gates can improve performance of SWAP-only routing algorithms.

Many other methods exist in addition to the ones above, such as that used by Google's Cirq \cite{CirqDocs}, or that proposed by Li et al. \cite{Li2018}, who also propose a technique for finding initial qubit placements. Research also exists regarding approaches that consider alternative factors when routing, such as differing qubit error rates \cite{Murali2019}, and even approaches that use quantum computers for quantum compilation \cite{Khatri2018}.

\section{Qubit Routing with Q-learning}

%% Qubit Routing with Q-learning

The RL formulation described in Section 2 was sufficient as a proof of principle, namely to prove that a modified Q-learning paradigm could be successfully applied to domains involving combinatorial action spaces, such as qubit routing. However, it was clear that a new formulation would be required in order to realistically tackle the problem of routing qubits, and to compete with the current state-of-the-art.

An illustration of the key issue is as follows. Under the formulation in Section 2, the system performs one layer of gates, one layer of swaps, and so on --- the core assumption is that gates occur instantaneously. This is not the case in reality, however --- quantum gates (e.g.\ a CNOT) take some time to perform, and quantum compilers will therefore endeavour to schedule swaps together with gates, potentially combining them in the same timestep.

Figure \ref{fig:new_rl_example} shows a circuit to be routed on a target topology (with given initial layout), and two possible solutions. Under the formulation in Section 2, the first CNOT gate would occur immediately, and one SWAP would be required to make the second CNOT possible. Figures \ref{fig:new_rl_1} and \ref{fig:new_rl_2} represent two different solutions that each require one ``action'' (in RL terms), and therefore appear equivalent, since they both use a single layer of SWAPs. However, in reality, $c_1$ can occur in two timesteps, since the first CNOT and the SWAP can occur in parallel, while $c_2$ must take three timesteps. $c_1$ is therefore the optimal solution, in terms of added circuit depth, but the RL formulation provides no way of telling that this is the optimal choice.

As this example demonstrates, an RL system could never hope to be optimal if it relies merely on implicit execution of logical quantum gates, since it could only strive to minimise SWAP depth in this case. We therefore propose a new RL formulation that explicitly allows for gates and swaps to be performed in the same timestep. The result is a system that realistically represents a quantum compiler routing pass, and can be fairly benchmarked against other state-of-the-art approaches. We also propose several other significant changes to the RL approach, which allow its performance to meet (and in some cases surpass) that of the state-of-the-art.

\subsection{Summary of the new formulation}

The key point of our proposed formulation is that the system needs to be able to ``mix in'' the swaps with the gates. The state-action formulation is thus as follows:

\begin{enumerate}
  \item The state now contains a set of ``protected nodes'' --- this contains nodes that are currently involved in a gate
  \item The initial state can contain protected nodes as well --- instead of just reducing the gates instantly as before, the system schedules them to be performed in the first timestep % changed
  \item The annealer can only select from swaps that don't involve any protected nodes --- it effectively ``fills'' the current timestep with some swaps (or none), while the currently scheduled gates are happening. % changed
  \item The action is performed by removing the gates that had previously been scheduled (and updating the qubit targets, if any), performing the swaps, and protecting nodes whose targets are now adjacent to them. % changed
\end{enumerate}

Effectively, this amounts to scheduling some gates, scheduling some swaps that don't conflict with them, and then updating the state in response to this action --- this means that gates and swaps are mixed into the same actions, but the gates are still \textit{mandatory}, i.e.\ gates are performed as soon as their two qubits land next to each other.

%TODO: define "performance"

\subsection{State representation}

Our preliminary investigations found that the ``nodes-to-target-nodes'' representation used in Herbert and Sengupta's RL approach \cite{Herbert2018} is too sparse, and therefore needs to be condensed into something that can be learned from more readily. The new representation (or feature selection function $\phi$) we propose here is a distance vector $\underline{\mathbf{d}}$, such that $\underline{\mathbf{d}}[i]$ represents the number of qubits that are a distance of $i$ from their targets. Besides a significant performance increase, another benefit of this representation is that it scales better --- rather than scaling with the number of qubits $n$, the state representation now scales with the diameter of the connectivity graph, which is $O(\sqrt{n})$ for a grid, and may be as little as $O(\log{n})$ in some cases \cite{Herbert2018a}.

The new representation is still injective, since many different scenarios can map onto the same distance vector, hindering the learning process. It is therefore helpful to allow the agent to distinguish situations in which its action choice will end up more or less constrained by the currently scheduled gates --- we add another component to our proposed feature selection function, $\underline{\mathbf{e}}$, such that $\underline{\mathbf{e}}[i]$ represents the number of nodes $n$ that have $i$ edges conforming to the following conditions:

\begin{enumerate}
  \item The edge neighbours $n$, and lies along the shortest path to $n$'s target
  \item The edge does not involve a currently protected node
\end{enumerate}

\subsection{Learning equation}

We found that assessing the quality of only individual states causes significant instability in the learning process, mainly due to poor action choices tarnishing the reputation of states that might actually be quite high-quality. Instead of assessing the quality of individual states, our new system assesses the quality of (state, next state) pairs. Like traditional Q-learning, it attempts to capture the idea of quality being assigned to a state and an action, but by using the next state instead of the action itself, the system avoids having to deal with the combinatorial action space.

The system now concatenates the state vectors for $s_t$ and $s_{t+1}$ and passes this new representation to the neural network. In other words, the system uses a new feature selection function $\Phi(s_t,s_{t+1}) = (\underline{\mathbf{d}}_{s_t},\underline{\mathbf{e}}_{s_t},\underline{\mathbf{d}}_{s_{t+1}},\underline{\mathbf{e}}_{s_{t+1}})$. Another key difference is the fact that annealing with the ``target'' model (using DDQN terminology) must now be conducted when learning from past experiences (``replay''). The new learning equation is thus as follows:

\begin{equation}
Q(s_t,s_{t+1}) = r_t + \gamma \max_{a+1}Q(s_{t+1},\text{env}(s_{t+1},a_{t+1}))
\end{equation}

A major advantage of this method is that it requires no static heuristics in order to train, and random actions are now more useful in training --- since quality is assigned to state \textit{transitions} rather than the states themselves, a poor action choice no longer tarnishes the quality of good states, improving stability. The training runtime is optimised by severely limiting the number of annealing steps in replay --- 10 steps is sufficient to reach maximum performance.

A fixed gate reward is issued for each pair of mutually-targeting qubits that are brought next to each other (i.e.\ when a gate is made possible). A distance reduction reward is also necessary in order to ensure that the system scales well to quantum architectures with a higher diameter --- this is just a fixed constant for each qubit that is brought closer to (but not next to) its target.

\section{Results}

%% Results

In this section, we evaluate our DQN system on a variety of quantum circuits and architectures, by comparing it to other routing algorithms in state-of-the-art compilers. The main compilers we benchmark against are CQC's \tket, IBM's Qiskit, and Google's Cirq. Other compilers exist, but Tan and Cong \cite{Tan2020} found \tket\ and Qiskit to be the leaders in the space, so this selection is sufficient to demonstrate how our approach compares to the industry standard.

The code for our DQN system is available on GitHub \cite{GitHub_Code_Repo}.

\subsection{Benchmarking setup}

When referring to the \textit{performance} of a given system, we mean how it behaves in terms of circuit depth overhead --- that is, better performance means lower circuit depth overhead. We instead use the term \textit{runtime} to refer to how long a given system takes to run.

We consider a SWAP gate as a primitive operation taking a single time step (same time as a CNOT) --- this assumption represents the simplest fair method of comparison, and has precedent in the literature \cite{Cowtan2019} (see Discussion for more on this). We have verified that the other compilers also output circuits with a mix of SWAPs and CNOTs, rather than performing SWAP decomposition. Throughout the following benchmarks, we have disabled every sort of compiler pass except for the routing process itself, in order to ensure that our comparison is fair and pertinent to the task at hand. This includes placement routines --- we have chosen to use random initial placements instead.

For the baseline systems, we downloaded the most recent versions compatible with Python 3.7. Where hyperparameters made a difference, we chose values that maximised performance --- in practice, we found that the number of trials for Qiskit's StochasticSwap algorithm was the only parameter that heavily impacted the results. For CQC's \tket, we disabled the use of BRIDGE gates, so it could be fairly compared to our SWAP-only algorithm --- we briefly assess the impact of such gates, as well as SWAP decomposition, in the Discussion section below.

\subsection{A word on runtime}

It is worth noting that our DQN system requires re-training for each target architecture, and potentially a round of hyperparameter optimisations, which critics may view as a drawback. However, in NISQ-era quantum computing, \textit{classical} (compiler) runtime is not such a concern, if \textit{quantum} runtime (CDO/CDR) can be improved at all. In addition, new quantum architectures are not developed every day, and once an RL agent (or ``model'') has been trained, it can be re-used on the same architecture indefinitely.

\subsection{Training}

In RL, it is common practice to train up a few different models and pick the best. For each of the following benchmarks, we trained a series of models for each architecture on randomly-generated sets of circuits, and the models were then run on separate test sets. In situations where there was significant variation in quality between models in training, only the best quality models were retained and subsequently used in testing. Such cases are clearly identified below, and we still run through the same total number of test circuits, for fairness.

\subsection{Single full-layer circuits}

\begin{figure}[h]
\centering
\begin{subfigure}[c]{.45\linewidth}
  \centering
  \includegraphics[height=\textwidth]{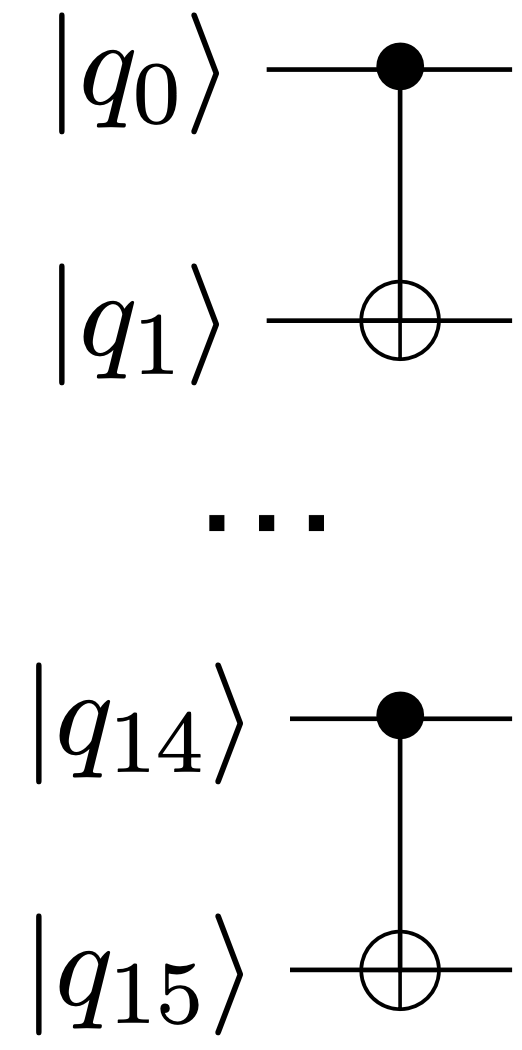}
  \caption{A single full-layer circuit with 16 qubits}
  \label{fig:single_full_16}
\end{subfigure}
\hspace{5mm}
\begin{subfigure}[c]{.45\linewidth}
  \centering
  \includegraphics[height=\linewidth]{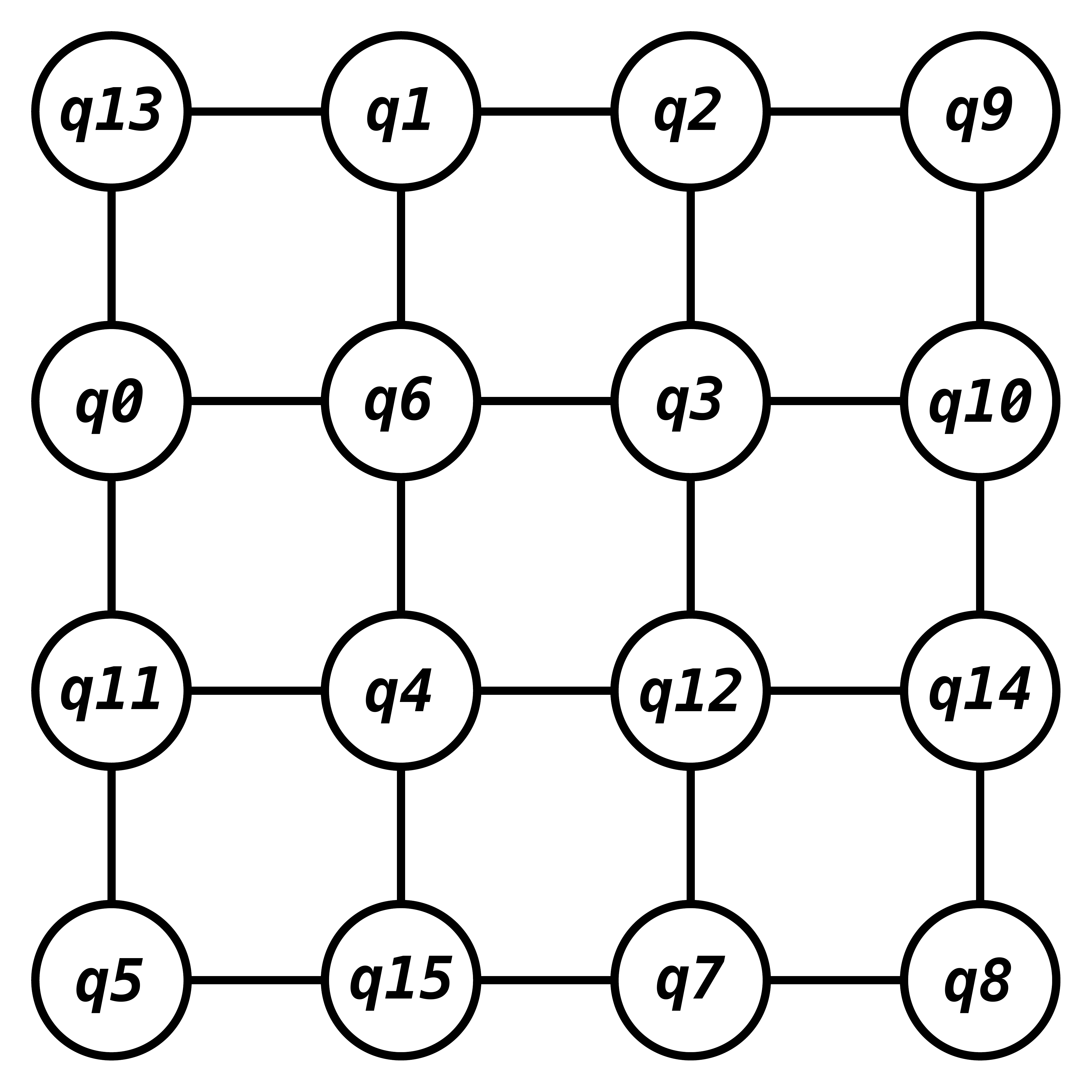}
  \caption{A 4x4 grid architecture with random initial placement}
  \label{fig:4x4_grid}
\end{subfigure}
\caption{An example of a 16-qubit single full-layer circuit, and a 4x4 grid architecture for it to be executed on}
\label{fig:full_layer_circuit_example}
\end{figure}

The first benchmark involves single full-layer circuits on increasing grid sizes. More precisely, these are n-qubit circuits, each with $\lfloor \frac{n}{2} \rfloor$ disjoint gates. This benchmark represents the worst kind of situation for a routing algorithm to deal with, since the original depth is very low ($d = 1$) but the number of gates to schedule is maximal for this depth. Figure \ref{fig:single_full_16} shows an example of such a circuit with 16 qubits.

Clearly, single full-layer circuits can be scheduled immediately on grid architectures if the correct initial placement is chosen, so a random placement is used in order to test the effectiveness of the routing scheme. Figure \ref{fig:4x4_grid} shows an example of a 4x4 grid architecture that could be used to execute the circuit represented by Figure \ref{fig:single_full_16}. For the random initial placement shown, only the gate between qubits $q_2$ and $q_3$ can be performed immediately --- SWAP gates must be inserted in order to make the other gates possible.

\begin{figure}
  \centering
  \includegraphics[width=\linewidth]{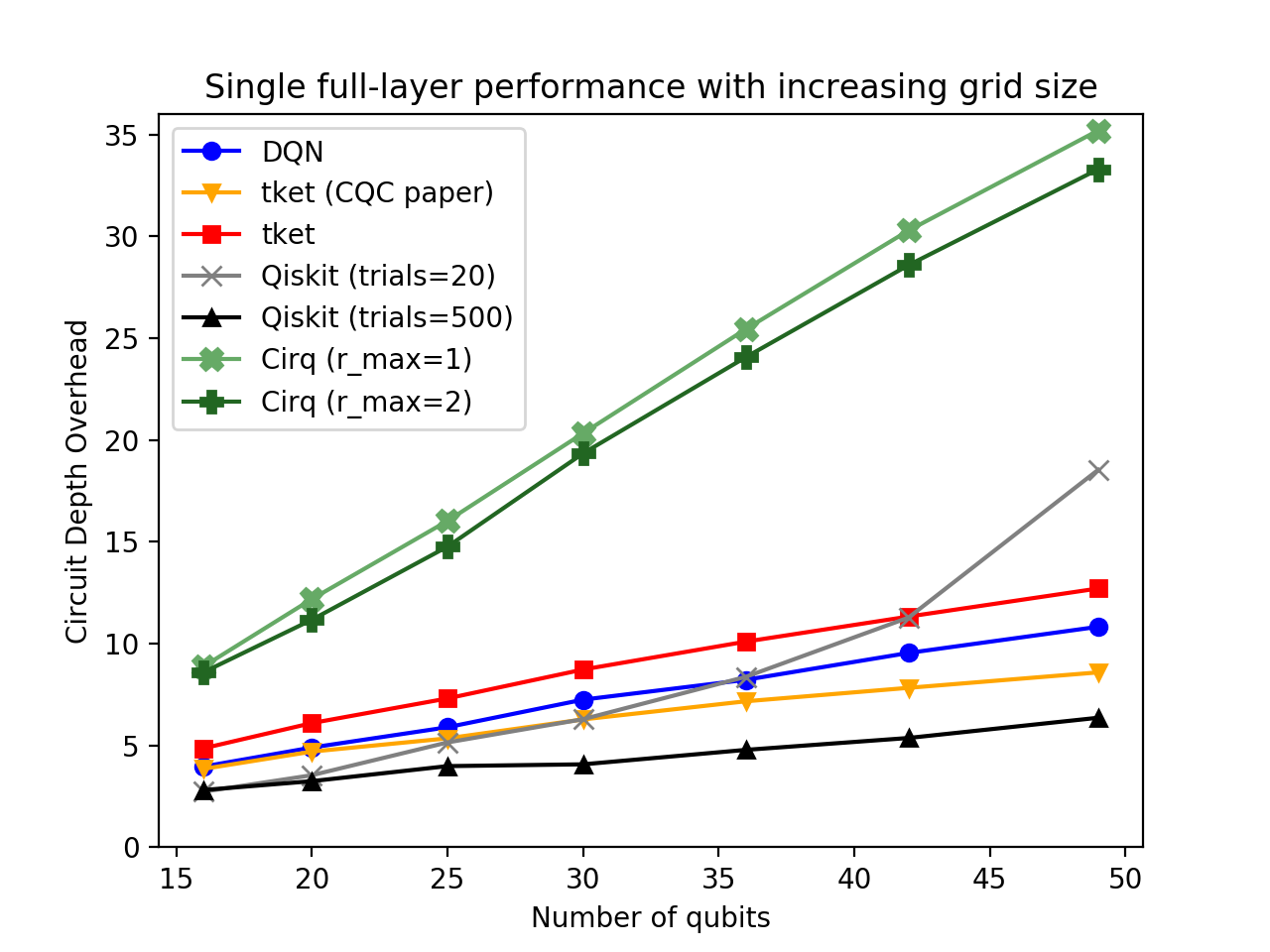}
  \caption{Single full-layer performance on various grid topologies}
  \label{fig:grid_scaling_benchmark}
\end{figure}

Figure \ref{fig:grid_scaling_benchmark} shows how CDO increases with increasing grid size. For each system, five batches of 100 test circuits were executed and the results were averaged. For the DQN system, the model was retrained on a separate training set also consisting of similar circuits, once per batch, for each different grid size, using a number of training circuits that increased linearly with the number of qubits.

Qiskit is clearly the best system for this benchmark, which is unsurprising, since it is the only system here that schedules all of the necessary swaps before scheduling all of the original (logical) gates at once, which is an effective strategy for maximally dense layers. However, the number of trials had to be increased significantly from the default value of 20 in order to reach this level of performance, which some might view as a drawback of the system.

We were unable to replicate the results in CQC's paper \cite{Cowtan2019}, so we include their reported results as well as the results we obtained from running \tket\ as described above. The DQN system outperforms the current version of \tket\ (although it is unable to outperform their reported results), and its performance scales sublinearly with the number of qubits, which is encouraging. Cirq's performance is poor, even with a better parameter value --- an $r_{max}$ of more than 3 rapidly becomes intractable, with minimal benefit. We therefore exclude Cirq from the benchmarks that follow.

\subsection{Multi-layer circuits}

\begin{figure}
\centering
\begin{subfigure}[c]{\linewidth}
  \centering
  \includegraphics[width=0.8\linewidth]{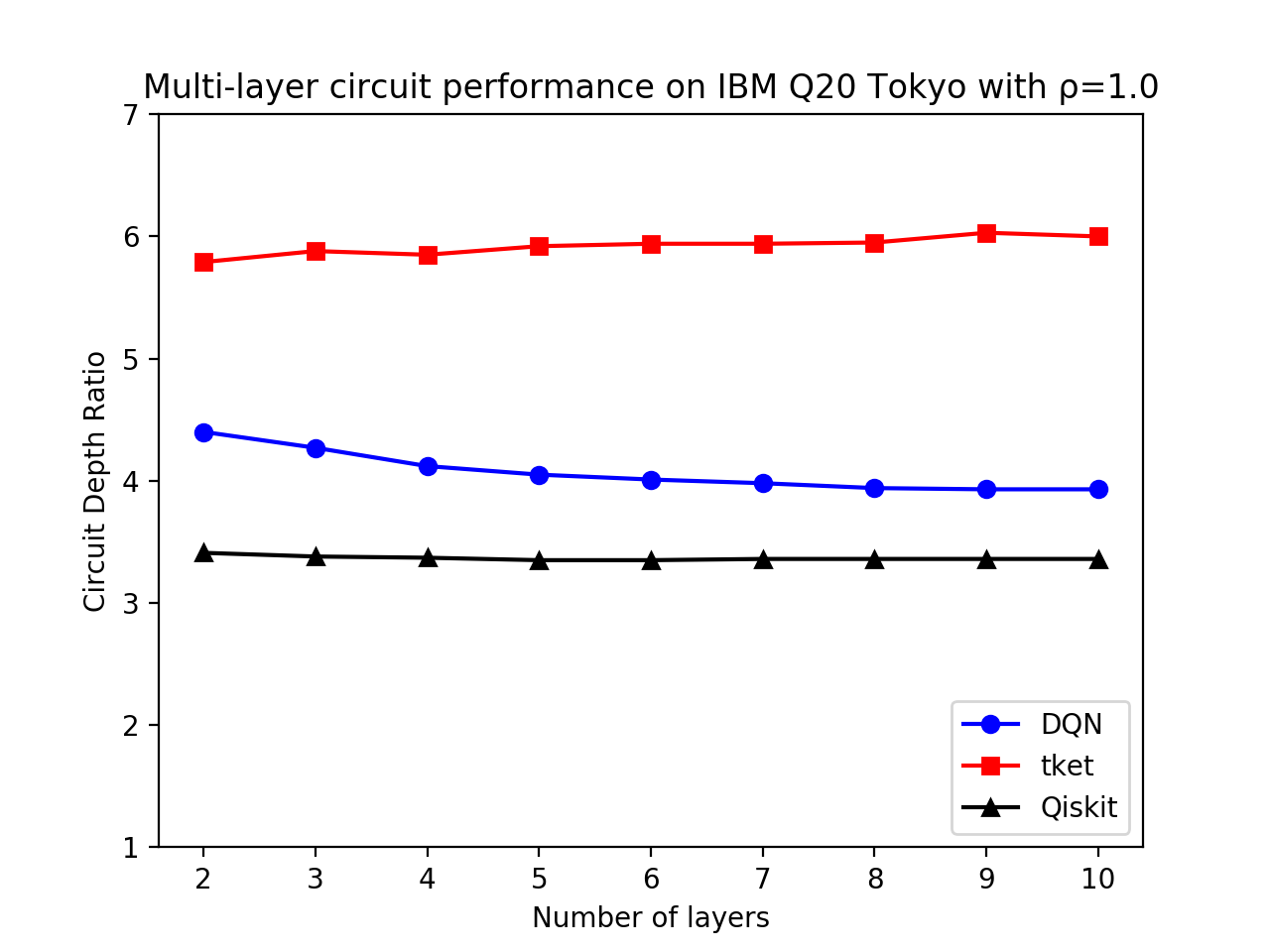}
  \caption{$\rho=1.0$}
  \label{fig:n_full_layer_1}
\end{subfigure}
\begin{subfigure}[c]{\linewidth}
  \centering
  \includegraphics[width=0.8\linewidth]{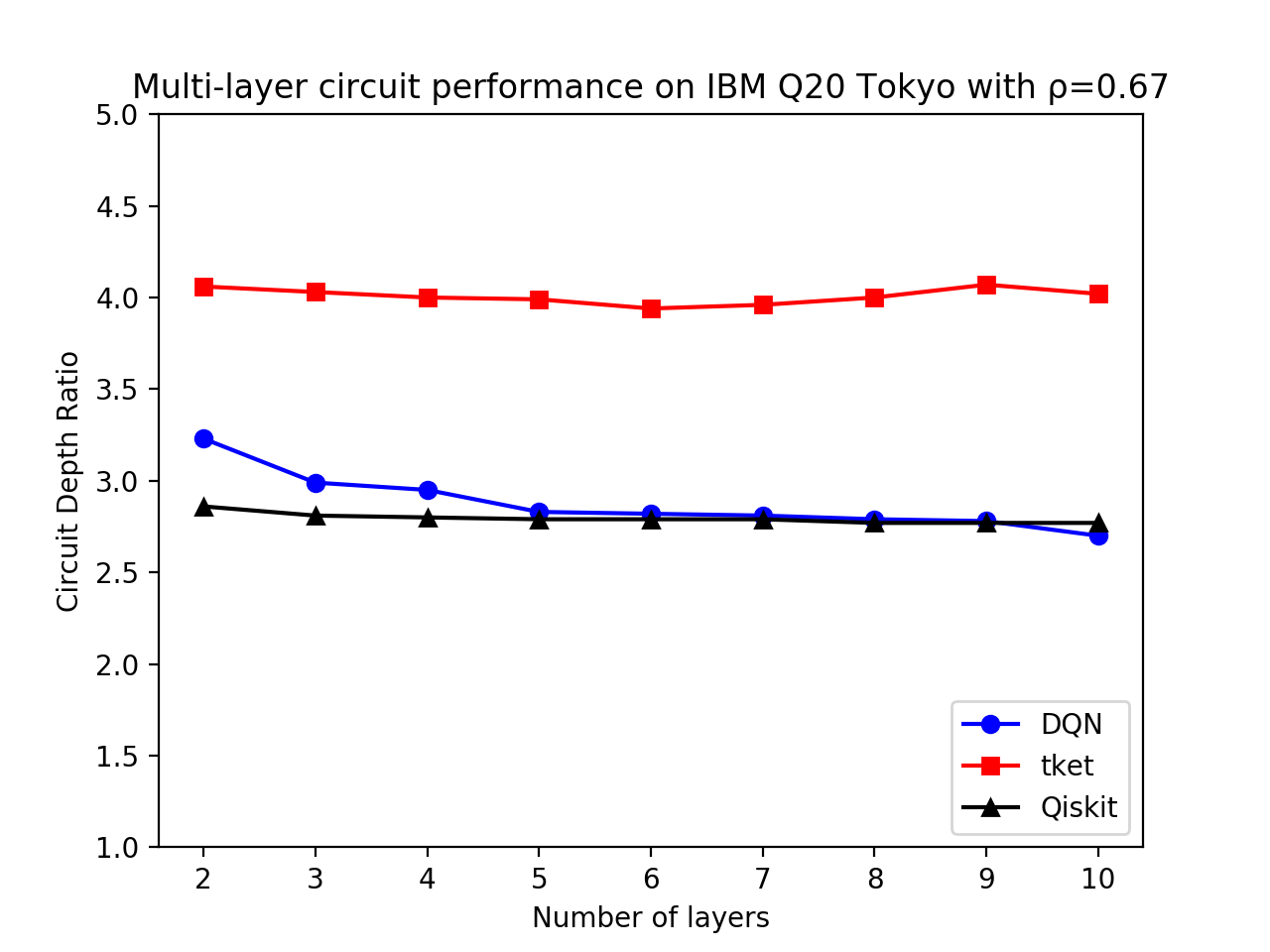}
  \caption{$\rho=0.67$}
  \label{fig:n_full_layer_67}
\end{subfigure}
\begin{subfigure}[c]{\linewidth}
  \centering
  \includegraphics[width=0.8\linewidth]{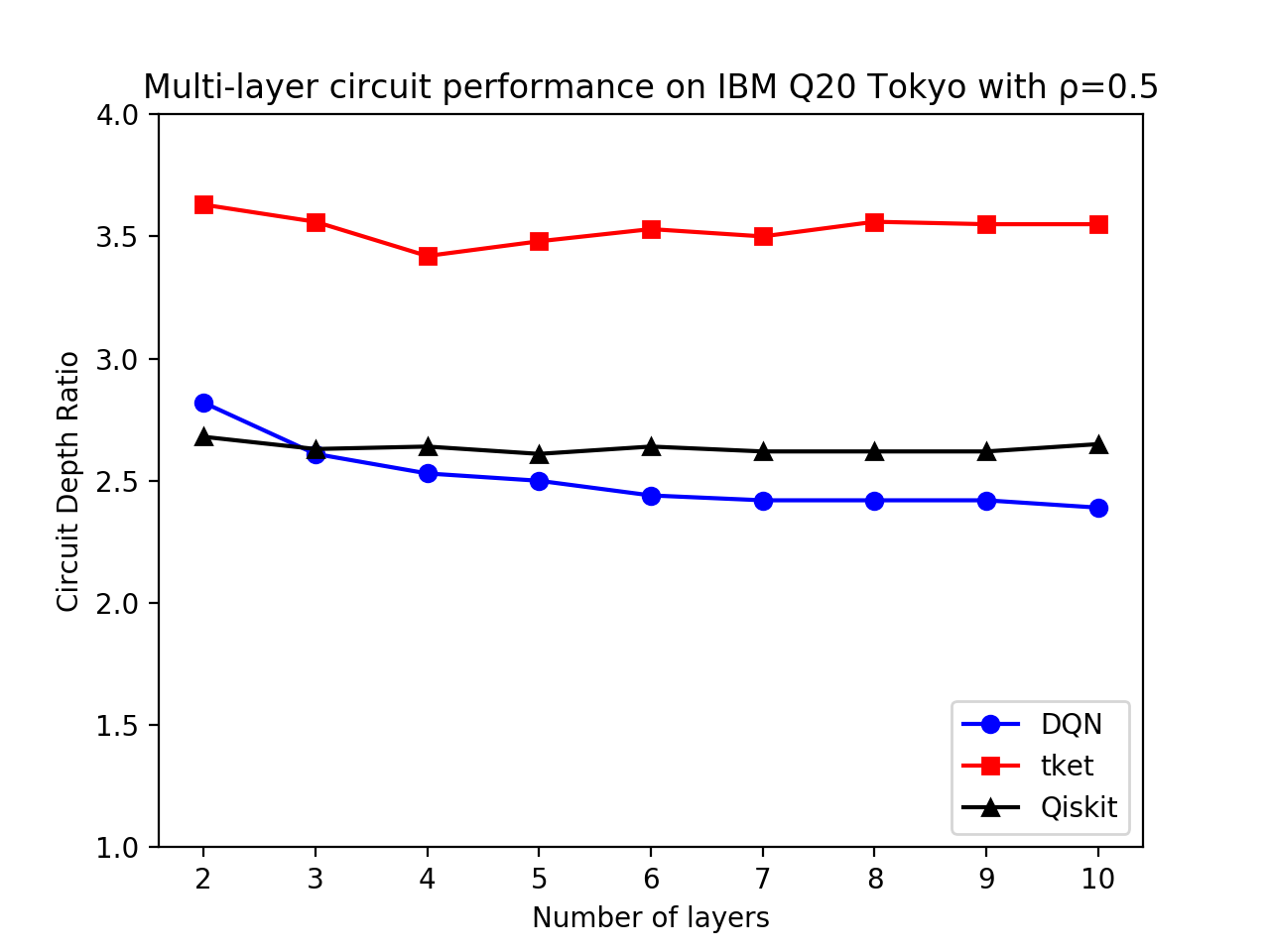}
  \caption{$\rho=0.5$}
  \label{fig:n_full_layer_5}
\end{subfigure}
\caption{Multi-Layer performance on the IBM Q20 Tokyo with differing circuit densities}
\label{fig:n_full_layer}
\end{figure}

\begin{figure*}
\begin{subfigure}{0.4\linewidth}
  \centering
  \includegraphics[width=\linewidth]{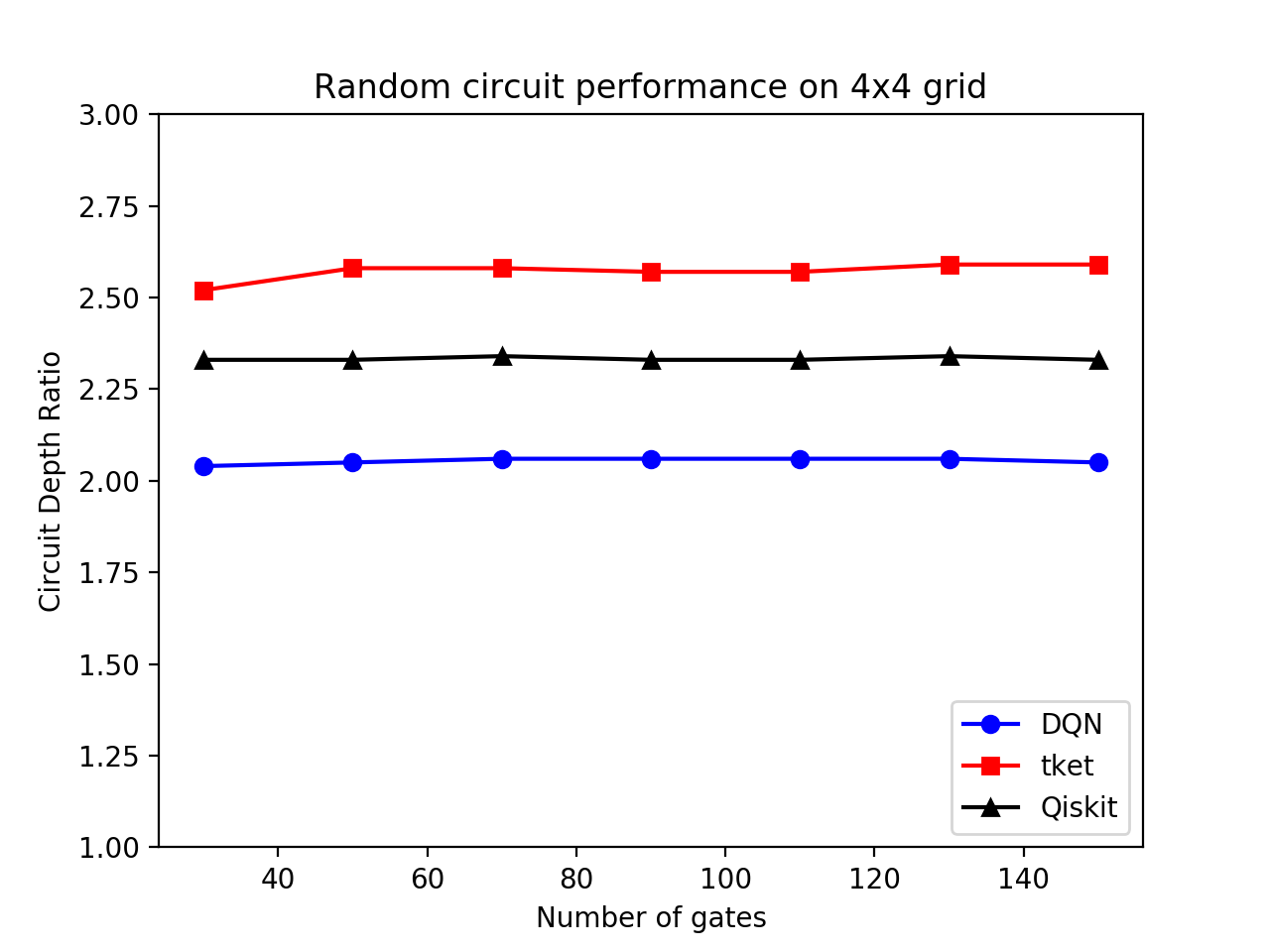}
  \caption{4x4 grid}
  \label{fig:random_circuits_benchmark_4x4}
\end{subfigure}
\begin{subfigure}{0.4\linewidth}
  \centering
  \includegraphics[width=\linewidth]{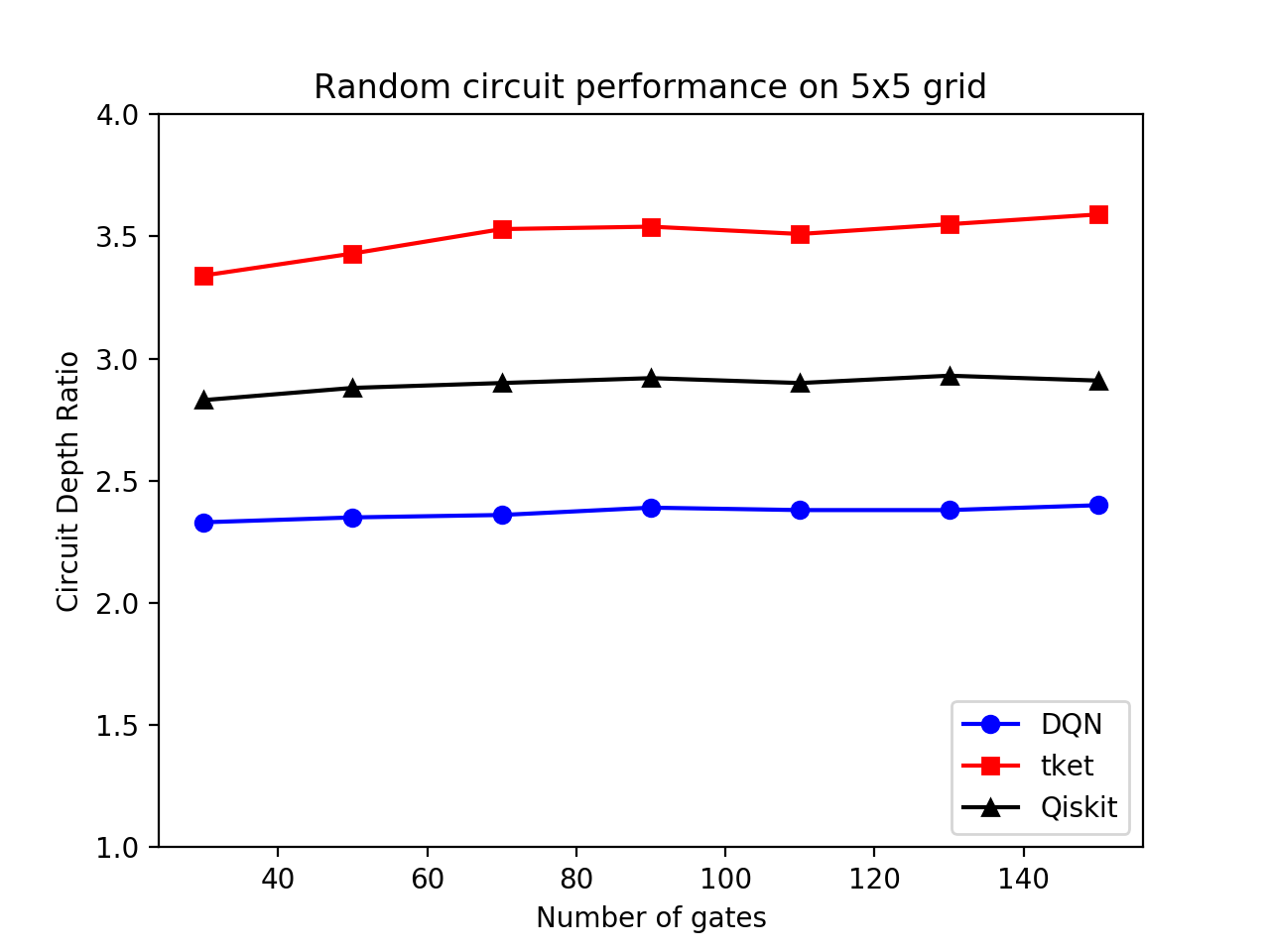}
  \caption{5x5 grid}
  \label{fig:random_circuits_benchmark_5x5}
\end{subfigure}
\begin{subfigure}{0.4\linewidth}
  \centering
  \includegraphics[width=\linewidth]{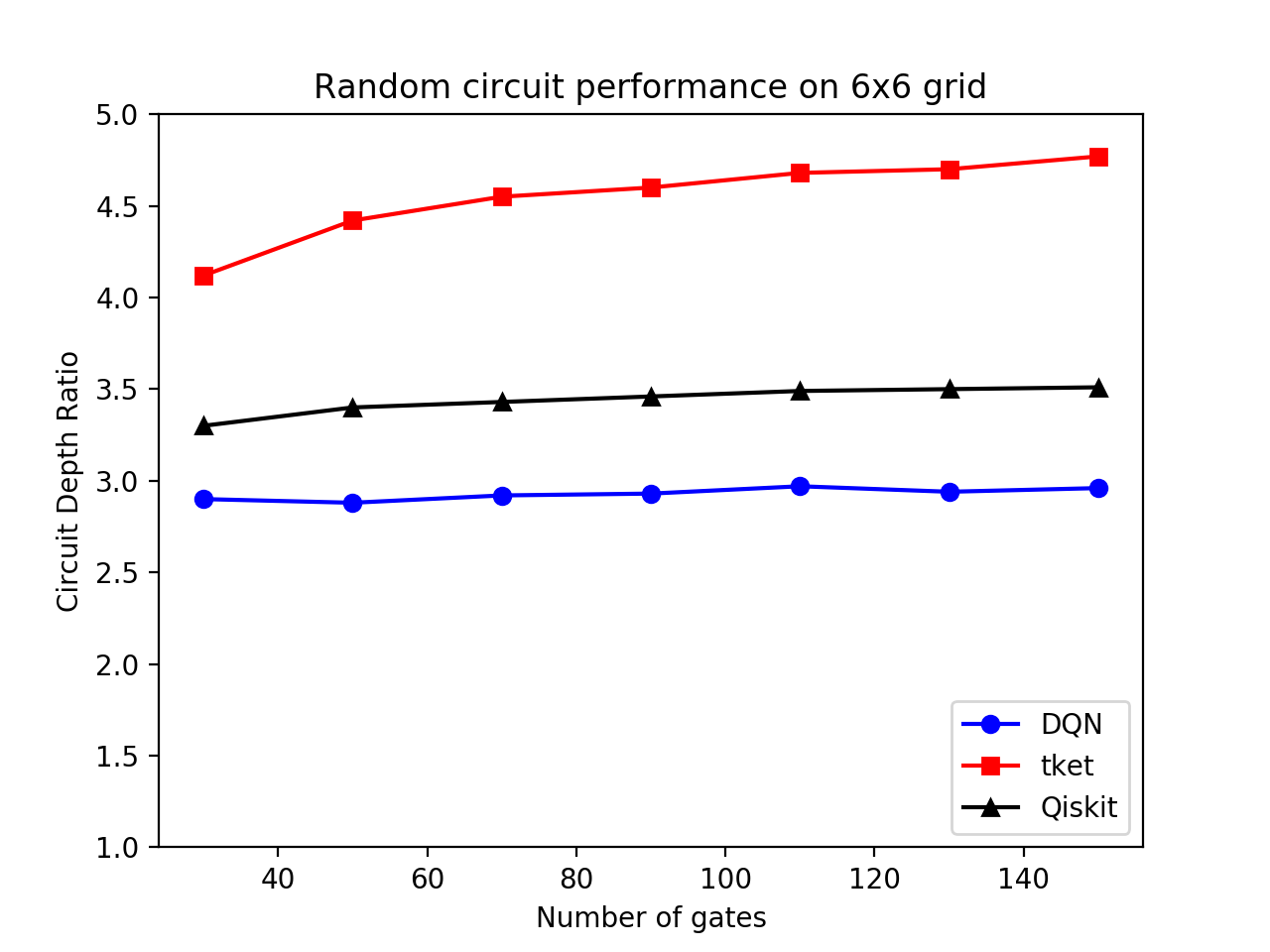}
  \caption{6x6 grid}
  \label{fig:random_circuits_benchmark_6x6}
\end{subfigure}
\begin{subfigure}{0.4\linewidth}
  \centering
  \includegraphics[width=\linewidth]{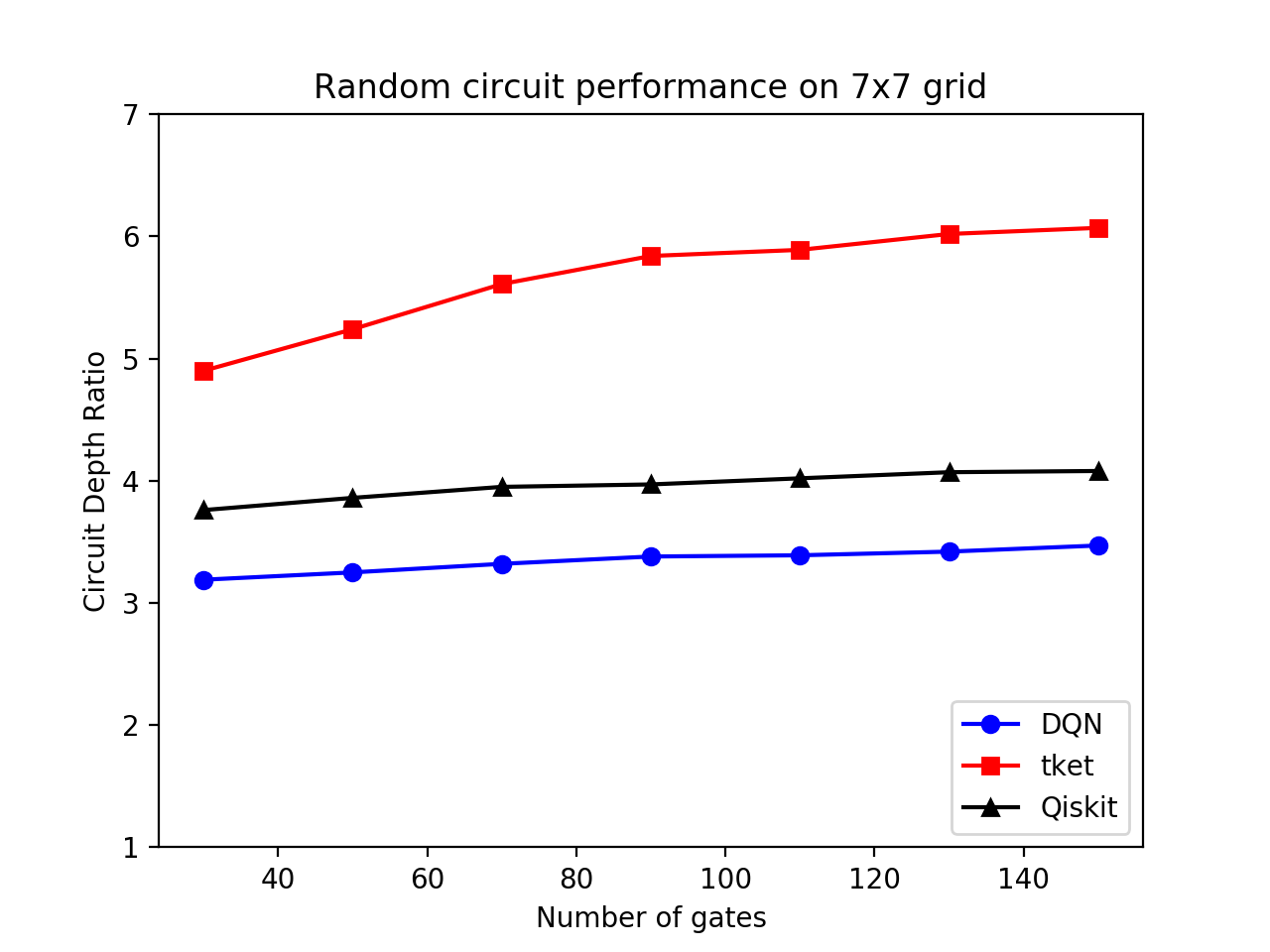}
  \caption{7x7 grid}
  \label{fig:random_circuits_benchmark_7x7}
\end{subfigure}
\caption{Random circuit performance on four quantum architectures}
\label{fig:random_circuits_benchmark}
\end{figure*}

Another important type of circuit to consider is the multi-layer circuit. These are circuits composed of a series of $N$ layers with density $\rho$, such that each layer will have $\rho \lfloor \frac{n}{2} \rfloor$ gates. A density of $\rho=1.0$ thus yields a series of full layers, and the number of gates in such a circuit is maximal for the given depth. Lower densities naturally lead to a less strict layer structure, reducing the number of gates in a circuit of given depth.

Figure \ref{fig:n_full_layer} shows the performance data for three circuit densities. The system was trained on 2-layer circuits with $\rho=1.0$, since we found that training on circuits with more layers actually worsened performance. The 3 best models out of 5 were selected for testing.

Qiskit exhibits good performance on full-layer circuits, since it employs its single-layer strategy for each (maximally dense) layer in sequence. Such behaviour is evident from its perfectly constant CDR in Figure \ref{fig:n_full_layer_1}. The DQN system also performs well in this case, outperforming \tket\ by about a third. It comes within about 20\% of Qiskit's performance on circuits with 10 layers.

As density is decreased, the performance of the DQN system begins to surpass that of Qiskit on deeper circuits. This is not a surprise, since Qiskit rigidly schedules each layer in sequence and ignores gates in future layers. DQN's performance improvement is a good sign as we turn towards random circuits below, which mostly have layer densities in the range $[0.25,0.45]$.

\subsection{Random circuits}

\begin{figure*}
\begin{subfigure}{0.4\linewidth}
  \centering
  \includegraphics[width=\linewidth]{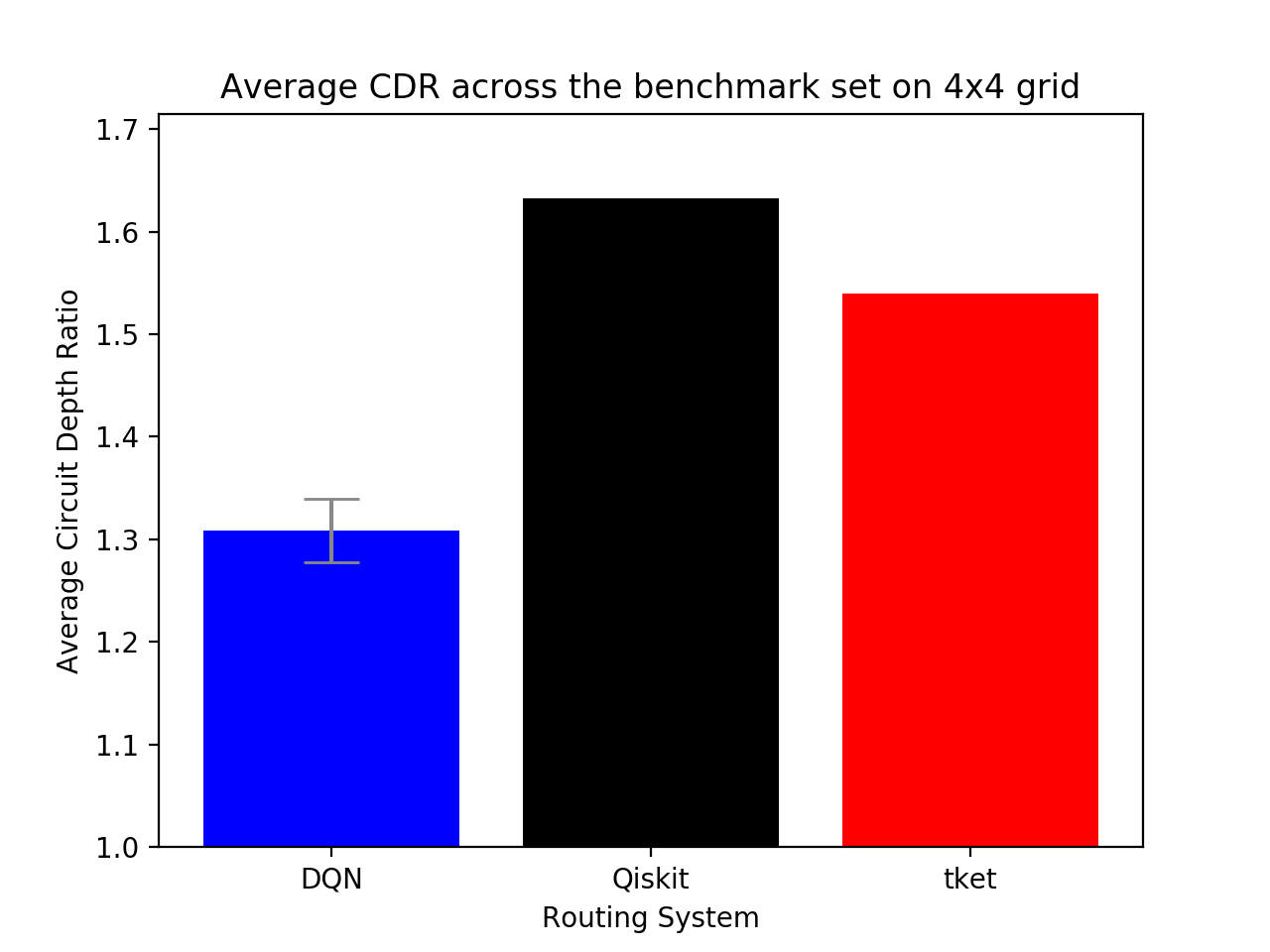}
  \caption{4x4 grid}
  \label{fig:realistic_test_set_benchmark_4x4}
\end{subfigure}
\begin{subfigure}{0.4\linewidth}
  \centering
  \includegraphics[width=\linewidth]{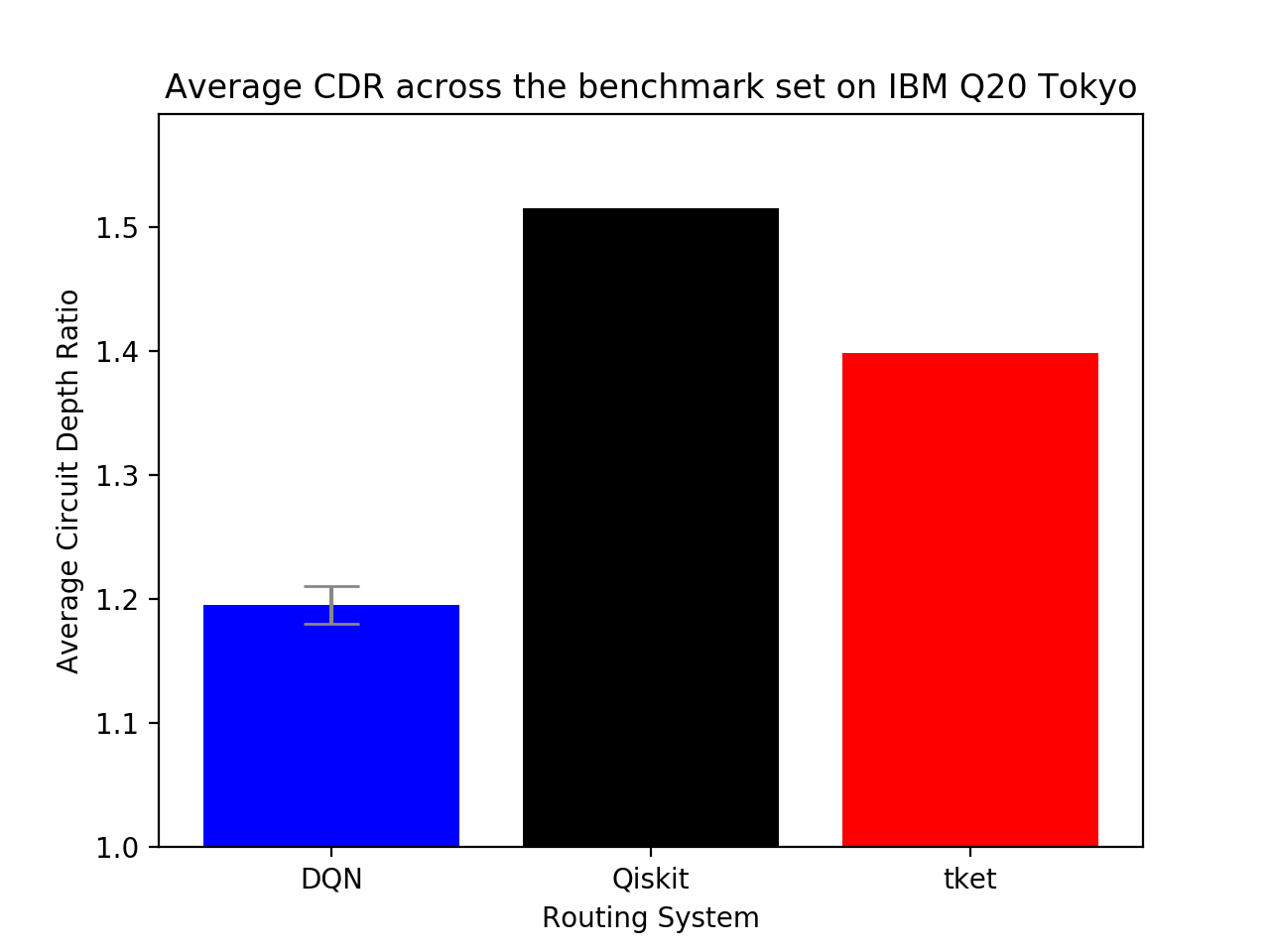}
  \caption{IBM Q20 Tokyo}
  \label{fig:realistic_test_set_benchmark_tokyo}
\end{subfigure}
\begin{subfigure}{0.4\linewidth}
  \centering
  \includegraphics[width=\linewidth]{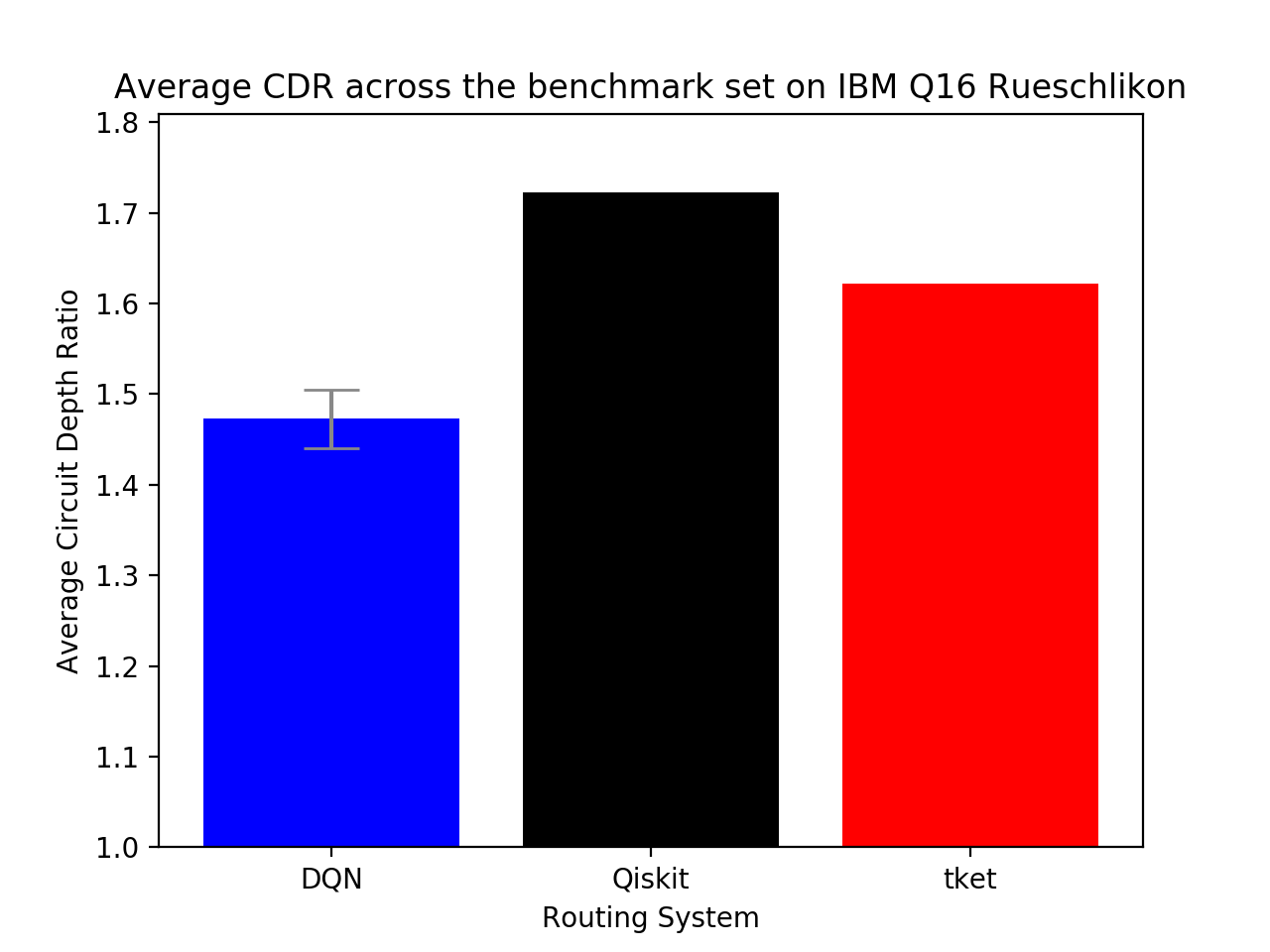}
  \caption{IBM Q16 R\"uschlikon (2x8 grid)}
  \label{fig:realistic_test_set_benchmark_rueschlikon}
\end{subfigure}
\begin{subfigure}{0.4\linewidth}
  \centering
  \includegraphics[width=\linewidth]{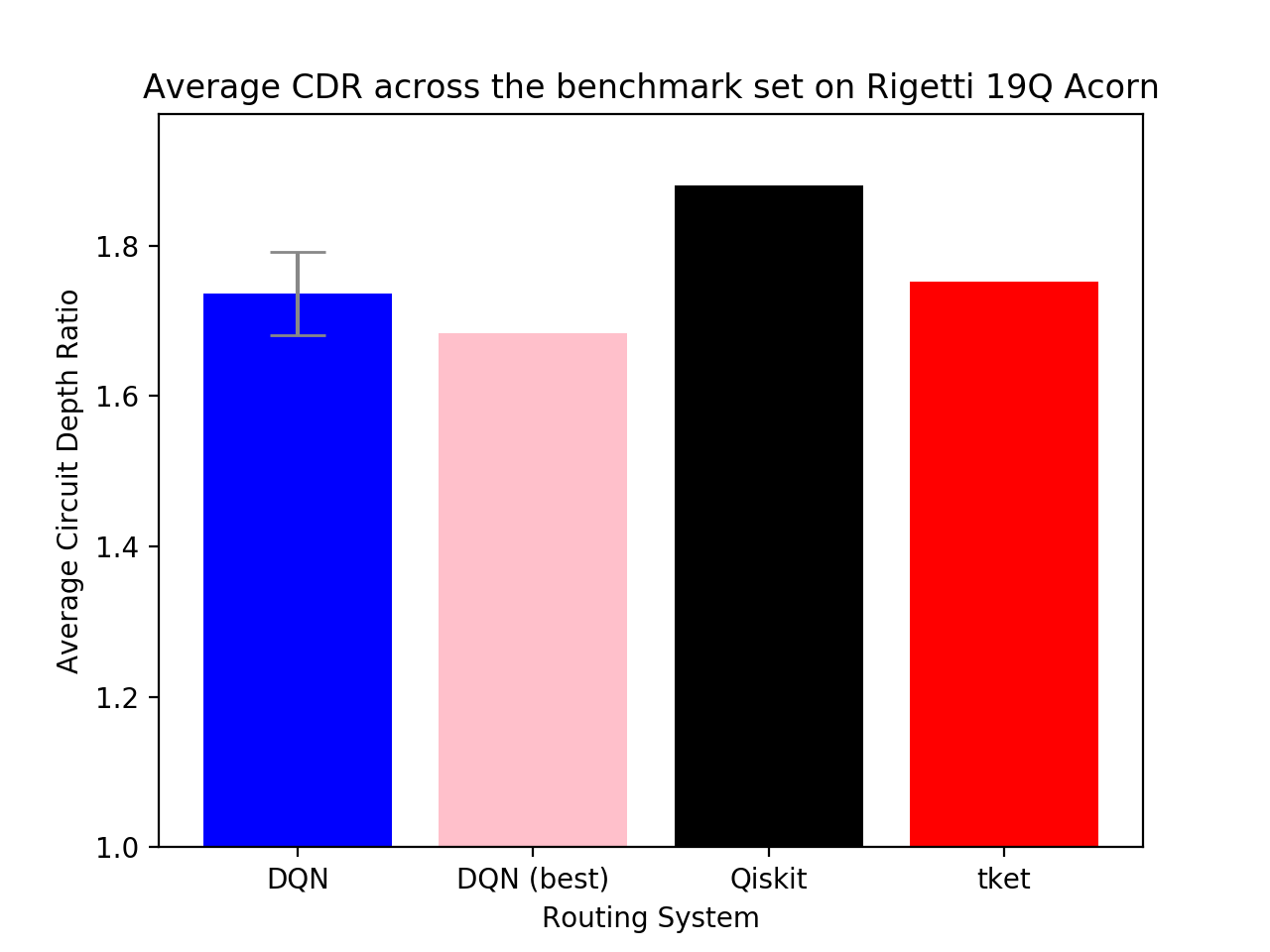}
  \caption{Rigetti 19Q Acorn}
  \label{fig:realistic_test_set_benchmark_rigetti}
\end{subfigure}
\caption{Realistic circuit performance on four quantum architectures}
\label{fig:realistic_test_set_benchmark}
\end{figure*}

Random circuits are a reasonable simulation of real quantum circuits. They are generated by adding gates between random qubits, leading to circuits with low layer densities. Figure \ref{fig:random_circuits_benchmark} shows the performance data for each system on four different quantum architectures. For each datapoint, the systems were executed on five batches of 100 test circuits each, and the results were averaged. For each architecture, we trained 16 DQN models with different hyperparameters on separate training sets consisting of random circuits, and retained the highest quality models for testing.

The DQN system has the best performance across all architectures and circuit sizes in the above plots, which is very encouraging. In particular, it is encouraging to see that the DQN system is able to maintain best-in-class performance on larger quantum architectures. For random circuits, average layer density increases with the number of gates in the circuit --- despite this, the DQN system's CDR still remains lowest on random circuits with 1000 gates, across all architectures.

As the grid size increases, hyperparameter optimisation becomes ever more important --- in particular, it is important to slow down the annealer's temperature decay, so that optimal actions can indeed be found. Nevertheless, these results demonstrate that the best DQN models are able to surpass the best competitor (Qiskit, here) on architectures of up to about 50 qubits, even as circuit depth increases, which suggests that the DQN system will remain competitive on most near-term quantum architectures and circuits.

\subsection{Realistic test set}

As a final benchmark, we sought to test each system on a set of real quantum circuits. We chose the test set of 158 circuits used by Zulehner \cite{Test_Set} and filtered out any circuits with a depth of 200 or more (due to runtime constraints). The final benchmark set thus consisted of 95 realistic quantum circuits, ranging from 3 to 16 qubits, depths of 5 to 199, and 5 to 240 CNOT gates.

We ran each system on four different realistic architectures. For each architecture, we trained 5 DQN models on random circuits with 50 gates, and generated five sets of random initial placements (one placement per circuit in each set). We then ran the system on the benchmark set, once per model and initial placement set, yielding a total of 25 runs through each circuit. We chose not to isolate the best models here, in order to give an indication of average-case performance. For the other systems, we repeated each circuit and initial placement five times with the same parameters. We also used a fixed seed when generating the placements, such that each system would use the same placement sets.

Figure \ref{fig:realistic_test_set_benchmark} shows the mean CDR for each system and quantum architecture. The error bars represent one standard deviation, obtained over the mean CDRs of each model. We only plotted error bars for the DQN system, since the error bars for the other systems were too small to plot --- for the former, variance between models is far more significant than variance between different runs of the same model, meaning that most of the variance between runs arises from variation in the quality of models.

The performance gap between the DQN system and the other baselines on the 4x4 grid and the IBM Q20 Tokyo is significant --- as the error bars illustrate, even the worst model on each architecture outperforms \tket\ by at least 11\% (and the best yields CDRs up to 17\% lower). The DQN models are also able to outperform the other baselines by a good margin on the IBM Q16 R\"uschlikon, which is effectively a 2x8 grid and thus has worse connectivity than a 4x4 grid (see next section for a discussion of architecture connectivities).

The DQN system runs into some issues on the Rigetti 19Q Acorn. Most notably, we found that training stability decreased, and that the performance gap between DQN and its competitors narrowed. Nonetheless, two models were somewhat higher quality than the other three --- isolating these two (see Figure \ref{fig:realistic_test_set_benchmark_rigetti}) yields a performance that is still better than that of \tket, albeit more marginally than on other architectures. The error bar was too small to plot here since the two models have very similar performance, but the benchmark did still run through each placement set five times, for consistency. In any case, with both models, every run through the test set yielded a lower CDR that \tket's average, so we can be reasonably confident that the DQN system is indeed still able to outperform the others on this architecture when training a few models and selecting the best (which is standard practice in RL).

\section{Discussion}

%% Discussion

Overall, the DQN system's performance throughout the benchmarks has been very positive. Performance on the layerised benchmarks was good, and the system still fared well against its competitors; looking instead at the random and real circuit benchmarks, the DQN system outperforms the other state-of-the-art baselines across all of the quantum architectures we tried. We would argue that such benchmarks (especially the realistic circuits) are the most important, since they most accurately represent the type of circuit that a routing algorithm may end up tackling in a real-world context, and it is therefore very encouraging to see how well the DQN system performs here.

Another key point to note is that the DQN approach is very flexible --- it has a wealth of hyperparameters that can be optimised for each specific architecture. On the other hand, the approaches used in state-of-the-art compilers tend to have very few hyperparameters and are somewhat fixed. In particular, Qiskit's strategy of focussing on each layer in sequence is wasteful on low-density circuits since not all qubits will be involved in a gate in a given layer, so swapping idle qubits is necessary to help schedule future gates. In fact, one can obtain a lower bound for the CDR that such a method could achieve on a given architecture by generating a random layer of gates (of a fixed target density) and initial placement, and computing half of the average furthest distance between any pair of qubits involved in a gate. This bounds the number of layers of SWAP gates required to schedule a given layer of logical gates. Doing so for a 7x7 grid, using the correct density for random circuits with 1000 gates, for example, yields a CDR that is still higher than that of the DQN system. This means that even with an infinite number of trials, Qiskit will not be able to outperform the DQN system for this architecture and circuit size. Such a result demonstrates the inherent limitation of treating layers separately when routing.

The DQN system struggles slightly on architectures with poor connectivity, notably the Rigetti 19Q Acorn. One possible explanation is the fact that the DQN system cannot distinguish between situations in which shortest paths cross and those that don't, and thus cannot predict upcoming conflicts. On such architectures, there are very few shortest paths between any two nodes (versus e.g.\ a grid), so choosing the path that minimises conflict is key. Another problem is when multiple qubits are all waiting to interact with the same one --- the system's state representation has no way of prioritising the movement of such qubits, despite the fact that their interactions are crucial for the progress of the routing process. This drawback is especially hard-hitting on architectures with poor connectivities, where its action choice is heavily constrained --- the system might get stuck in local minima since it can't tell which qubit is the source of the bottleneck.

Ultimately, it is not such a problem if the system performs sub-optimally on architectures with poor connectivities, since we can reasonably assume that connectivity will only improve in future. However, the above points could help motivate future work on the system, especially with respect to very large grid sizes. While the DQN system's performance was still best-in-class on the grid sizes we tried (which are sufficiently large to indicate near-term performance), breaking ties between shortest paths and tackling the wider issue of qubit priority will likely be essential to unlock better performance on even larger grid architectures, which is especially relevant as we move towards architectures with ever increasing numbers of qubits in the future.

\subsection{Applying SWAP decomposition}

Throughout the Results section, we considered both CNOT and SWAP gates to take one timestep each. This was the simplest fair method of comparing routing procedures (with precedent in the literature \cite{Cowtan2019}), since it is the most architecture-agnostic --- for example, for some quantum technologies, pulse-level optimisation can lead to SWAP gates executing in 1.5 timesteps (by using $\sqrt{\text{iSWAP}}$ gates) \cite{Gokhale2020}, and in future we can expect a wider variety of such optimisations to emerge. However, at the moment, SWAP gates must be performed via decomposition into CNOT gates, and they thus take three timesteps to execute (as illustrated in Figure \ref{fig:swap_gate_cnot}). While the DQN system has not been optimised with this in mind, we nonetheless sought to assess each routing system's performance when performing such decomposition.

\begin{figure}
\centering
\begin{subfigure}[c]{\linewidth}
  \centering
  \includegraphics[width=\linewidth]{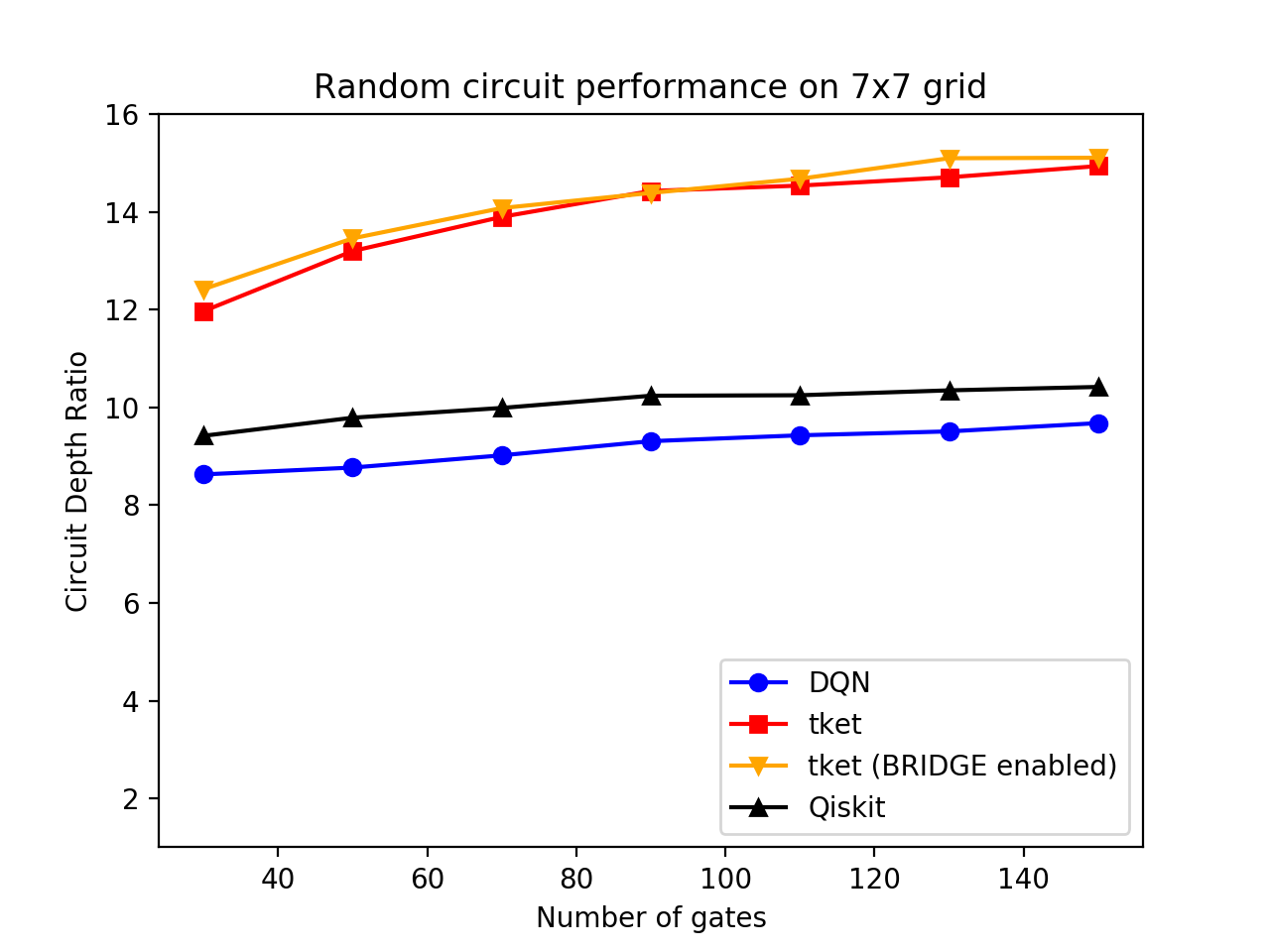}
  \caption{Random circuit performance on a 7x7 grid}
  \label{fig:random_circuits_decomposed}
\end{subfigure}
\begin{subfigure}[c]{\linewidth}
  \centering
  \includegraphics[width=\linewidth]{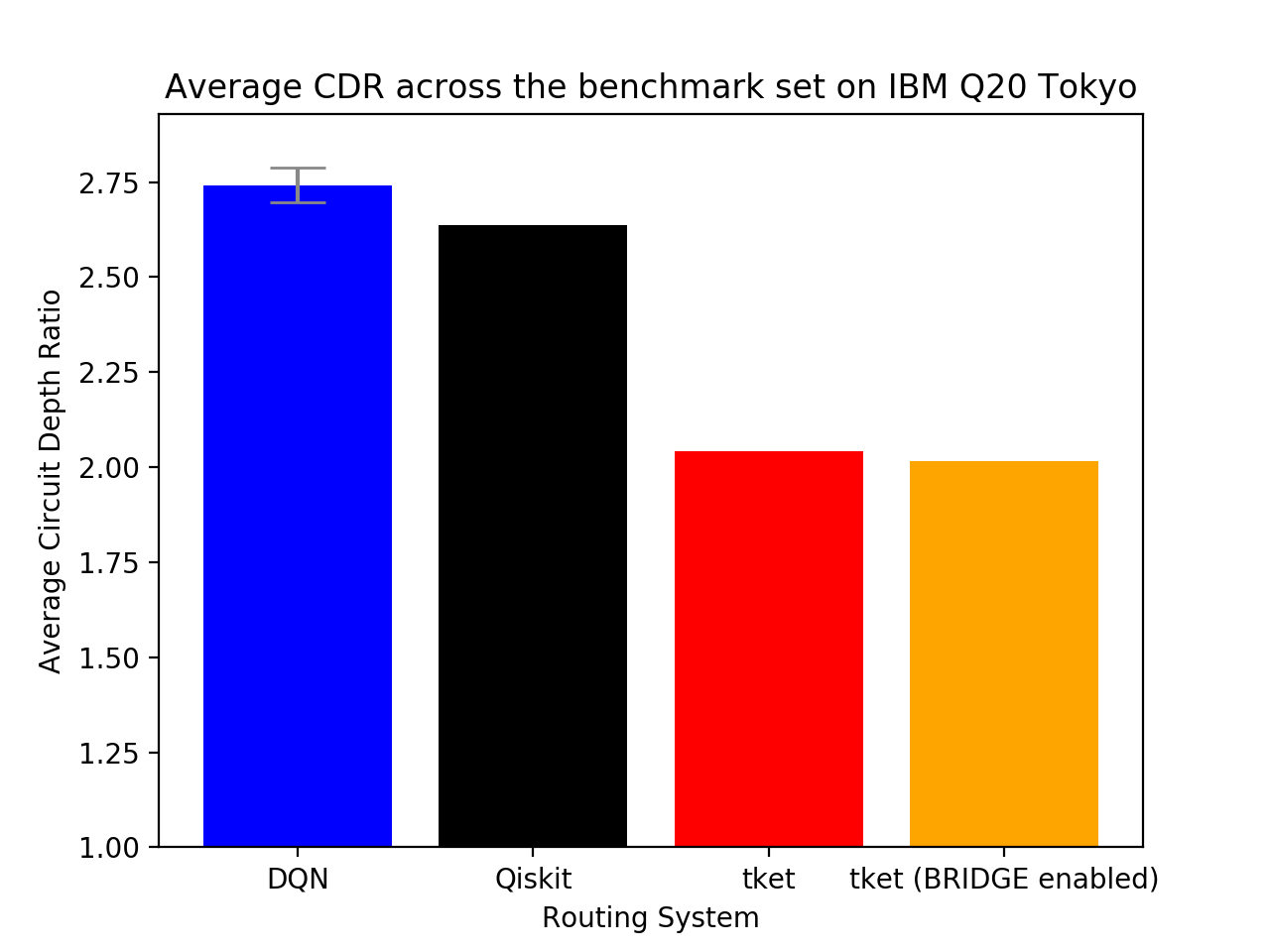}
  \caption{Realistic circuit performance on the IBM Q20 Tokyo}
  \label{fig:realistic_circuits_decomposed}
\end{subfigure}
\caption{Performance of each routing system, calculating CDR after SWAP decomposition into CNOT gates}
\label{fig:n_full_layer}
\end{figure}

Figure \ref{fig:random_circuits_decomposed} shows each system's performance on random circuits after performing SWAP decomposition, while Figure \ref{fig:realistic_circuits_decomposed} shows performance on realistic circuits. We also tried enabling BRIDGE gates for \tket, in both cases, and show these results separately. Encouragingly, the DQN system still outperforms its competitors on random circuits, even on such a large grid size --- in fact, the general shape of the graph remains largely unchanged. However, the ordering of the systems is reversed for realistic circuits. In particular, \tket\ is by far the best system when decomposing SWAPs into CNOTs, suggesting that its routing process might be optimised for this particular scenario. The relatively poor performance of DQN here (despite excellent performance when not performing decomposition) could be explained by a variety of factors, perhaps most significantly by the fact that it makes no effort to minimise SWAP count, only depth. When considering SWAPs and CNOTs to take the same amount of time, performing a redundant SWAP carries no penalty in the RL formulation --- however, when performing decomposition, an unnecessary depth cost may additionally be incurred from such redundancy, especially on sparse (e.g. realistic) circuits with low CDRs. Furthermore, the DQN system has no way of optimising its action choice specifically with decomposition in mind --- awareness that a SWAP takes three timesteps would allow the system to choose SWAPs that allow for ``pipelining'', that is, beginning a SWAP while another is ongoing, in order to minimise depth overhead.

To reiterate, DQN's poor performance relative to \tket\ and Qiskit here is not surprising, since in this work, the DQN system was not optimised with SWAP decompositon in mind. However, the RL formulation can certainly be modified to mitigate both of the above issues --- much in the same way that this paper proposes mixing SWAPs and CNOTs into the same timesteps, future work may well allow SWAP gates to be scheduled with an awareness of their decomposition, thus allowing them to occur ``out of time'', enabling decompositions that yield a lower depth overhead than otherwise possible. BRIDGE gates may also be added as potential actions for the annealer to choose from, and models can be trained on realistic circuits in order to cope with common patterns that arise therein (rather than training still on random circuits, as we do above). We firmly believe that with such improvements, a DQN system will still be able to remain competitive in the most realistic scenarios. Meanwhile, a more static method such as Qiskit's presents no such opportunities for improvement, and its performance is therefore bounded.

Overall however, DQN's best-in-class performance on random circuits, even when performing SWAP decomposition, demonstrates that the system is well ahead of its competitors when considering routing as an isolated problem. We therefore believe that there are more gains to be had by adopting an RL approach in quantum compilers more generally, and certainly by implementing the improvements we outline above.

\subsection{Other recommendations for future development}

The feature selection function is the component that we have found most impactful throughout the work --- we thoroughly believe that with a more rich state representation, the DQN agent will be able to achieve even higher performance in complex scenarios. The main problem with the current representation is the fact that a lot of information is lost when converting the full state into a mere distance vector. The information about available swaps helps somewhat, but in practice this does little to help break ties when qubits are far away for their targets. A new representation might encode some information about shortest paths between mutually-targeting qubits and their potential for conflict, as well as information about which qubits should be prioritised. Ultimately, there is a balance here between the size of the representation and the information it is able to capture. It is also possible that a massive network with massive amounts of training could learn its own optimal representation, in the true spirit of ``deep'' learning --- we would be curious to see whether this works. Furthermore, many RL methods employ some form of lookahead, in which a series of future episodes are simulated in order to choose the best action --- this could help the DQN system to predict upcoming bottlenecks and react accordingly.

Another possible improvement relates to automating the learning process. At the moment, choice of the number of training episodes is somewhat arbitrary, but it would certainly be more useful (and reliable) to employ a deterministic scheme with some well-formed criteria, such as detecting when the weights of the neural network have converged.

Furthermore, the system currently lacks the ability to use BRIDGE gates, or delay gates instead of scheduling them as soon as possible. Extending the RL paradigm to incorporate such characteristics is certainly achievable, and can simply be done by adding these as possible actions to the RL formulation. It is also worth noting that \tket\ not only uses BRIDGE gates, but it also has optimisation passes to simplify chains of CNOT gates --- while we disabled such functionality in order to perform a fair comparison, it would certainly improve \tket's performance when considering circuit depth after SWAP decomposition, a necessary step for current architectures. Adding similar functionality to the DQN system would be a simple yet important task, in order to remain competitive as a full compilation process (i.e.\ beyond mere routing).

\subsection{Another word on runtime}

At the moment, the system is not very well-optimised in terms of runtime --- we have always preferred to run the system longer, or use a more exhaustive method, in order to minimise CDO and CDR. Indeed, optimising the system's runtime could be an entire paper in itself, which is why we have shied away from directly comparing the runtime of our system to that of the other baselines. That said, it is important to give at least some indication of the timescales involved. For the realistic test set on a 4x4 grid, disregarding training time, the DQN system took about 2400 seconds to complete one run (of 100 circuits), while Qiskit (StochasticSwap) took about 36s and \tket\ took about 6s. The DQN system is clearly much slower, but for perspective, Qiskit's LookaheadSwap routing method \cite{LookaheadSwap} (which came second in the Qiskit Developer Challenge) is almost 4 times as slow as the DQN system on the 4x4 realistic circuits benchmark, despite only having as good a CDR as Qiskit's faster \lstinline{StochasticSwap} method. Equally, it is worth noting that newer CPU architectures with SIMD extensions (such as \lstinline{AVX} or \lstinline{FMA}) and GPUs could help improve the runtime of our method, without touching the code itself.

Besides, the runtime of the DQN system can certainly be reduced while still maintaining good performance. One such area for improvement is the annealer --- an adaptive scheme with a variable number of iterations would help greatly. In fact, other combinatorial optimisation techniques could be used, such as random restart hill-climbing \cite{Russell2010}. Another area for optimisation is clearly the size of the neural network used. In practice we found such changes to make little difference, but it is perfectly possible that the layer structure used at the moment is wasteful --- a more principled search would be necessary for each quantum architecture. Once again, this would be a worthy time investment, since new quantum architectures are developed infrequently --- the time required to develop a new architecture is clearly far greater than the time required for such a search. Equally, returning to a single state approach (with some scalability improvements) may prove effective while greatly diminishing training times due to the lack of annealing when replaying past experiences. Better parallelisation would also be useful --- the literature on this topic includes several methods that could be of use in an RL context \cite{Nair2015,Mnih2016,Clemente2017}.

\section{Conclusion}

%% Conclusion

In this paper, we have presented a Reinforcement Learning (RL) approach to address the problem of routing qubits on near-term quantum architectures. Our main contribution has been to improve the approach proposed by Herbert and Sengupta \cite{Herbert2018} so that it represents a realistic solution to the qubit routing problem, as would be expected from a quantum compiler, which then allowed us to benchmark the system against state-of-the-art approaches. We also made several other improvements to the original system, such as a new state representation and learning paradigm, which helped to improve its performance significantly.

The key research question throughout has been: can a DQN approach be used to perform qubit routing in quantum compilers, and if so, is it able to compete with state-of-the-art approaches? We would say the answer is emphatically, yes. The results demonstrate that a DQN approach is able to surpass the performance of other industry-standard approaches in realistic near-term scenarios, with a level of adaptability that is not possible with other more static approaches, which will be particularly useful as a wider variety of quantum architectures appear in future. Further work is required in order to maintain best-in-class performance when performing SWAP decomposition, but we are confident that this will be achievable with some modest improvements to the system. Overall, our work demonstrates the value of using an RL approach in the compilation of quantum circuits, and we hope that such an approach can bring further benefits in the space in future.

%%
%% The acknowledgments section is defined using the "acks" environment
%% (and NOT an unnumbered section). This ensures the proper
%% identification of the section in the article metadata, and the
%% consistent spelling of the heading.
\begin{acks}
Special thanks to Silas Dilkes from Cambridge Quantum Computing (CQC) for providing guidance on how to set up \tket\ for our benchmarks.
\end{acks}

%%
%% The next two lines define the bibliography style to be used, and
%% the bibliography file.
\bibliographystyle{ACM-Reference-Format}
\bibliography{publication}

\end{document}